\documentclass[sigconf, screen]{acmart}

\usepackage{xspace} 

\usepackage{amsthm}
\usepackage{amsmath}  
\usepackage{bm}

\def\eqref#1{equation~\ref{#1}}

\def\1{\bm{1}}

\DeclareMathAlphabet{\mathsfit}{\encodingdefault}{\sfdefault}{m}{sl}
\SetMathAlphabet{\mathsfit}{bold}{\encodingdefault}{\sfdefault}{bx}{n}

 \usepackage{hyperref}
\usepackage{amsmath}
\usepackage{graphicx}
\usepackage{multirow}
\usepackage{adjustbox}
\usepackage{tabularx}
\usepackage{tablefootnote}
\usepackage{booktabs}
\usepackage{array}
\usepackage{balance}
\usepackage{multirow}
\usepackage[normalem]{ulem}
\usepackage{color}
\definecolor{lightgray}{RGB}{215,215,215}
\usepackage{colortbl}  
\usepackage{xcolor}
\useunder{\uline}{\ul}{}
\usepackage{subfigure}
\usepackage{algorithm}  
\usepackage{algorithmicx}
\usepackage{amsfonts}
\usepackage[noend]{algpseudocode}
\usepackage{xcolor} 
\usepackage[inline]{enumitem}
\usepackage{tabularx}
\usepackage[utf8]{inputenc}
\usepackage[english]{babel}
\usepackage{amsthm}
\usepackage{bm}
\usepackage{acronym}
\usepackage{mdframed} 
\usepackage{lipsum}   
\usepackage{url}
\usepackage{booktabs}
\usepackage{multirow}
\usepackage{amsmath,amsthm}
\usepackage{amsmath}

\author{Jujia Zhao}
\email{zhao.jujia.0913@gmail.com}
\affiliation{
\institution{Leiden University}
\city{Leiden}
\country{Netherlands}
}
\author{Zihan Wang*}
\email{zhw.cypher@gmail.com}
\affiliation{
\institution{CISPA Helmholtz Center for Information Security}
\city{Saarbrücken}
\country{Germany}
}
\author{Shuaiqun Pan}
\email{s.pan@liacs.leidenuniv.nl}
\affiliation{
\institution{Leiden University}
\city{Leiden}
\country{Netherlands}
}

\author{Suzan Verberne}
\email{s.verberne@liacs.leidenuniv.nl}
\affiliation{
\institution{Leiden University}
\city{Leiden}
\country{Netherlands}
}

\author{Zhaochun Ren*}
\email{z.ren@liacs.leidenuniv.nl}
\affiliation{
\institution{Leiden University}
\city{Leiden}
\country{Netherlands}
}

\copyrightyear{2026}
\acmYear{2026}
\setcopyright{cc}
\setcctype{by}
\acmConference[SIGIR '26]{Proceedings of the 49th International ACM SIGIR Conference on Research and Development in Information Retrieval}{July 20--24, 2026}{Melbourne, VIC, Australia}
\acmBooktitle{Proceedings of the 49th International ACM SIGIR Conference on Research and Development in Information Retrieval (SIGIR '26), July 20--24, 2026, Melbourne, VIC, Australia}
\acmDOI{10.1145/3805712.3809719}
\acmISBN{979-8-4007-2599-9/2026/07}

\title[Unifying Search and Recommendation in LLMs via Gradient Multi-Subspace Tuning]{Unifying Search and Recommendation in LLMs via \\Gradient Multi-Subspace Tuning}

\begin{document}

\begin{abstract}
Search and recommendation (S\&R) are two integral components of modern online platforms, both aiming to model and satisfy user information needs. This shared objective motivates a unified modeling paradigm that enables richer user modeling and improves the effectiveness of both tasks.
Recent attempts to unify S\&R formulate item ranking in both tasks as conditional generation.
While this paradigm is promising, existing methods rely on full fine-tuning, which is computationally expensive and limits scalability. 
Parameter-efficient fine-tuning (PEFT) offers a more practical alternative but faces two critical challenges in unifying S\&R: 
(1) gradient conflicts across tasks due to divergent optimization objectives, and (2) shifts in user intent understanding caused by overfitting to fine-tuning data, which distort general-domain knowledge and weaken LLM reasoning.
To address these issues, we propose Gradient Multi-Subspace Tuning (GEMS), a novel framework that unifies S\&R with LLMs while alleviating gradient conflicts and preserving general-domain knowledge. 
GEMS introduces 
(1) \textbf{Multi-Subspace Decomposition}, which disentangles shared and task-specific optimization signals into complementary low-rank subspaces, thereby reducing destructive gradient interference, and 
(2) \textbf{Null-Space Projection}, which constrains parameter updates to a subspace orthogonal to the general-domain knowledge space, mitigating shifts in user intent understanding.
Extensive experiments on benchmark datasets show that GEMS consistently outperforms the state-of-the-art baselines across both search and recommendation tasks, and the gains remain consistent when scaling to billion-parameter LLMs.\footnote{Our code is available at \url{https://github.com/Polaris-JZ/GEMS}.}
\end{abstract}

\ccsdesc[500]{Information systems~Retrieval models and ranking}
\ccsdesc[500]{Information systems~Recommender systems}
\keywords{Unifying Search and Recommendation, Generative Recommendation, Multi-task Learning, Large Language Models}
\maketitle

\let\thefootnote\relax\footnotetext{$^*$Corresponding author.}
\section{Introduction}
\label{sec:introduction}
Recommender systems and search engines are two integral parts of modern online service platforms, both fundamentally aiming to model and satisfy user information needs~\cite{zhang2024towards,wu2024generative,zhang2026model,xu2026unveiling}.
Their shared objective reveals a natural connection, while their complementary user behaviors offer opportunities for mutual enhancement~\cite{shi2024unisar}.
Search queries capture users’ short-term intents that can inform timely recommendations, whereas recommendation models encode long-term preferences that can improve the personalization of search results~\cite{xie2024unifiedssr,lin2026verifiable}.
Integrating the two within a unified framework therefore represents a promising direction, enabling richer user modeling and potentially improving the overall effectiveness of both tasks~\cite{yao2021user}.

\begin{figure}[t]  
\setlength{\abovecaptionskip}{0cm}
\setlength{\belowcaptionskip}{-0.5cm}
    \centering    
    \includegraphics[width=\linewidth]{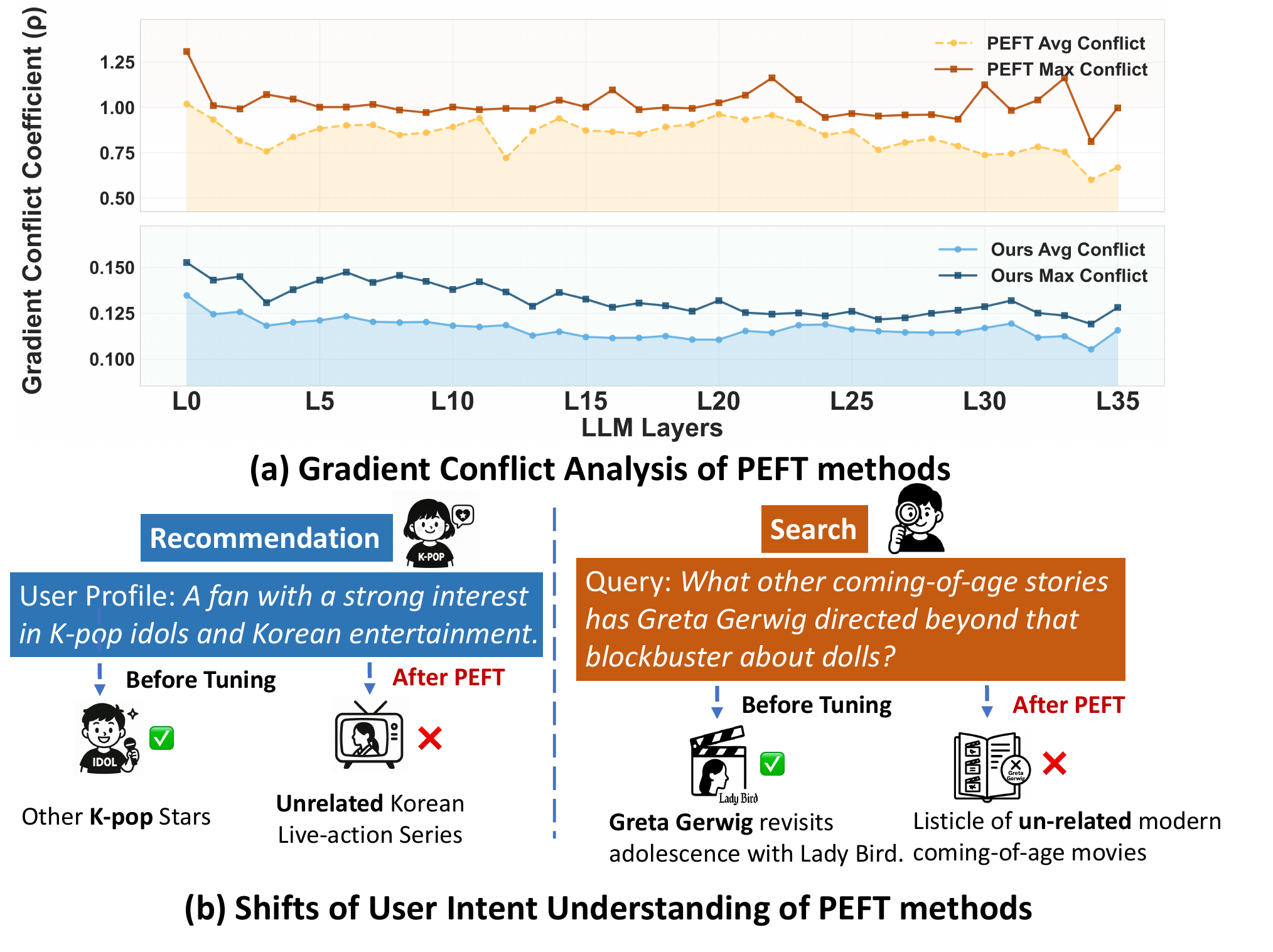}
    \caption{(a) Gradient conflict analysis across the layers of Qwen-3B under PEFT (e.g., LoRA) and our methods on the Qilin dataset. The Gradient Conflict Coefficient (\textbf{$\rho$})~\cite{du2018adapting} quantifies the degree of opposition between the gradients of the S\&R tasks (see Section~\ref{subsubsec:GCA}); lower values indicate less conflict. The results show that our method consistently achieves lower gradient conflicts compared to PEFT. (b) Shifts in user intent understanding in real-world S\&R.}
    \label{fig:intro}
\end{figure}

Early attempts to unify S\&R adopt shared transformer modules to process input features (e.g., user interaction histories or queries) and use task-specific heads to predict relevance scores between users and candidate items, followed by a ranking step to identify the target item~\cite{shi2024unisar,yao2021user}.
Although these methods show promise, they require considerable manual effort in designing model architectures and task-specific heads, which limits their scalability and practicality. 
In addition, the absence of end-to-end optimization makes them prone to local optima~\cite{li2024learning}.
Recent advances instead employ generative models that reframe ranking as conditional generation, using pre-trained language models (PLMs) to directly generate target item identifiers conditioned on user queries or interaction histories~\cite{penha2024bridging}.
These approaches are readily adaptable to different PLM backbones and support end-to-end training, offering higher flexibility and a more favorable optimization landscape than earlier paradigms.
Furthermore, generative methods naturally inherit the strong semantic understanding and reasoning capabilities of PLMs, enabling richer modeling of both user intent and item representations~\cite{shi2025unified}.

However, existing generative methods rely on full fine-tuning, which limits generalization and scalability and leads to substantial computational and memory costs when applied to prevailing large language models (LLMs).
To address these limitations, recent studies explore parameter-efficient fine-tuning (PEFT) techniques that update only a small subset of parameters (e.g., low-rank adapters or prefix tokens) while keeping most backbone weights frozen, thereby reducing the resource demands of large-scale adaptation~\cite{hu2022lora}.
Despite these advantages, PEFT presents two critical challenges when applied to unified S\&R tasks:
(1) \textbf{Gradient conflicts across tasks.} As shown in Figure~\ref{fig:intro}(a), significant disagreement between task gradients is observed across multiple LLM layers.
This arises because search and recommendation pursue inherently different objectives: search focuses on modeling query–item relevance within the current query context, whereas recommendation emphasizes capturing long-term user preferences from historical interactions.
These conflicting optimization signals lead to contradictory gradient directions, resulting in unstable training and degraded performance for both tasks.
(2) \textbf{Shifts in user intent understanding}.
During PEFT-based adaptation, overfitting to limited fine-tuning data often induces a semantic distribution shift in the representation space~\cite{fang2024alphaedit}.
This shift disrupts general-domain knowledge and weakens the model’s intrinsic language understanding and reasoning abilities to interpret user intent.
Consequently, the model may generate incorrect or inconsistent outputs even for queries it previously handled correctly (see Figure~\ref{fig:intro}(b)).

To address abovementioned challenges, we introduce a \textbf{G}radi\textbf{e}nt \textbf{M}ulti-\textbf{S}ubspace Tuning (GEMS) framework that unifies S\&R within LLMs 
while mitigating gradient conflict and preserving general-domain knowledge. 
GEMS is built on the core idea of \textbf{subspace tuning}: rather than updating all parameters along raw gradient directions, it projects gradients into a set of dominant low-rank subspaces learned from gradient statistics for optimization~\cite{zhao2024galore}.
The resulting updates are then projected back into the full parameter space, substantially reducing training-time memory overhead.
Compared with PEFT methods, subspace tuning eliminates the restrictive low-rank assumptions imposed by reparameterization~\cite{xia2024assessing} and remains deployment-friendly, as it introduces no additional adapter weights.


Building on subspace tuning, GEMS introduces two key components to tackle the above challenges:
(1) \textbf{Multi-subspace decomposition} further organizes the updates into three complementary subspaces:
a shared subspace capturing information consistently useful to both tasks, 
and two task-specific subspaces that encode signals unique to S\&R, respectively.
By explicitly disentangling shared and task-specific signals, GEMS mitigates gradient conflicts: 
updates along directions agreed upon by both tasks are routed to the shared subspace, while potentially conflicting signals are confined to their respective task-specific subspaces. 
Since these task-specific subspaces are constructed from inherently different gradient patterns, they exhibit minimal geometric overlap, substantially reducing the chance of destructive gradient interference when their updates are combined.
An adaptive gating mechanism further balances their contributions over training based on task dynamics, improving optimization stability.
(2) \textbf{Null-space projection} constrains updates to be orthogonal to the general-domain knowledge space, thereby preserving the model’s inherent language understanding and reasoning capabilities.
Specifically, GEMS estimates the principal representation space of the backbone model using general-domain pre-trained data and projects the combined gradient onto the null space of this knowledge space, effectively limiting representational drift and maintaining robust user intent understanding.
To evaluate the effectiveness and efficiency of GEMS, we conduct extensive experiments on benchmark datasets spanning both search and recommendation scenarios.
The results show that GEMS consistently outperforms strong baselines, including state-of-the-art S\&R models and leading PEFT approaches.

Our main contributions are summarized as follows:
\begin{enumerate*}[label=(\arabic*), leftmargin=*]
    \item To the best of our knowledge, ours is the first study to adapt LLMs with billions of parameters for unified S\&R without full fine-tuning, enabling efficient parameter updates and knowledge preservation.
    \item We introduce multi-subspace decomposition, which disentangles shared and task-specific optimization signals to mitigate gradient conflicts by separating consistent and conflicting gradient directions into complementary low-rank subspaces.  
    \item We develop null-space projection, which constrains parameter updates to be orthogonal to the general-domain knowledge space, reducing representational drift and preserving the model’s language understanding and reasoning capabilities.  
    \item We conduct extensive experiments across multiple benchmark datasets and LLM backbones, demonstrating that GEMS outperforms competitive baselines in S\&R performance, training efficiency, mitigation of gradient conflicts, and preservation of general-domain knowledge.
\end{enumerate*}
\section{Related Work}
\label{sec:related_work}

\noindent\textbf{Unifying search and recommendation.}
Unifying search and recommendation (S\&R) within a single model promises richer user modeling and mutual gains, and has therefore attracted growing attention~\cite{penha2024bridging,zhao2025unifying,zhang2024unified,zhao2022joint}.
Early attempts share transformer modules while attaching task-specific heads that score user–item relevance, followed by a final ranking step~\cite{yao2021user,zamani2018joint}.
For example, 
UnifiedSSR~\cite{xie2024unifiedssr} jointly models user behavior history in S\&R scenarios using a parameter-sharing dual-branch network and an intent-oriented session module.
UniSAR~\cite{shi2024unisar} models fine-grained user behavior transitions between S\&R through extraction, alignment, and fusion.
While effective, these methods demand substantial manual design and lack end-to-end optimization, making them vulnerable to local optima~\cite{li2024learning}.
More recent efforts employ generative models, leveraging PLMs to directly generate target item identifiers conditioned on user queries or interaction histories~\cite{penha2025semantic}.
Specifically, BSR~\cite{penha2024bridging} jointly trains generative models for S\&R tasks using atomic item identifiers.
GenSAR~\cite{shi2025unified} unifies generative S\&R by designing dual-purpose semantic and collaborative item identifiers.
While current generative methods implements greater practicality and end-to-end optimization, they rely on full fine-tuning, which faces computational and memory costs when applying to LLMs.

\noindent\textbf{Multi-task learning in LLMs.}
Although not yet widely applied to unifying S\&R, LLMs have emerged as powerful backbones for multi-task learning due to their strong contextual reasoning and generalization abilities~\cite{chung2024scaling,raffel2020exploring,wei2022emergent,wang2025graph,wang2025cooperative}. 
By leveraging instruction-based formulations and shared semantic representations, LLMs can jointly learn multiple related tasks, enhancing overall performance~\cite{wang2022super}.
Existing approaches can be grouped into two main directions. 
(1) Unified instruction-based fine-tuning, which fully shares model parameters across tasks and formats each task as a natural language instruction~\cite{shengyu2023instruction,huang2024unifit,wei2021finetuned,lyu2024macpo}. 
For instance, T0~\cite{sanh2021multitask} reformulates diverse datasets into prompted forms and fine-tunes a single model, achieving strong generalization to unseen tasks. 
While effective, this paradigm demands careful balancing of heterogeneous data and suffers from task interference when objectives conflict~\cite{ding2023mitigating}.
(2) Parameter-efficient multi-task learning, which introduces lightweight modules or experts for each task while sharing a common backbone~\cite{pfeiffer2020adapterfusion,feng2024mixture,yang2025mtl}. 
For instance, LoRA-MoE~\cite{dou2023loramoe} enhances multi-task learning by integrating a plugin Mixture-of-Experts architecture with specialized LoRA experts, utilizing localized balancing constraints to dynamically route and dedicate expert groups for distinct downstream tasks and knowledge retention.
While these modular approaches mitigate gradient conflicts, they introduce additional complexity and cost (e.g. routing mechanisms and multiple modules), and their performance hinges on effective assignment of tasks to the right experts~\cite{fedus2022switch,lepikhin2020gshard}.
In contrast, our proposed GEMS framework is based on subspace tuning, requiring no additional routing or auxiliary modules.
Additionally, GEMS learns task-specifc subspaces derived from task statistics, effectively capturing unique optimization directions for each task.

In this work, we efficiently unify S\&R within LLMs. 
The most closely related studies include~\cite{penha2024bridging,shi2025unified,penha2025semantic}.
However, they face two major challenges: 
(i) gradient conflict across tasks; and 
(ii) shifts in user-intent understanding. 
Our proposed GEMS, which
incorporates multi-subspace decomposition and null-space projection to alleviate gradient conflicts and preserve accurate user-intent understanding.

\section{Preliminaries}
\label{sec:pre}

\noindent\textbf{Unifying S\&R.}
We formulate the task of unifying S\&R as learning a single model that integrates users’ historical S\&R interactions to perform either task according to user needs.
In the search setting, the objective is to retrieve documents relevant to a user’s query while leveraging their historical S\&R interactions. 
In the recommendation setting, the objective is to suggest items based on the same historical interactions.
Formally, let $\mathcal{U}$ and $\mathcal{I}$ denote the sets of users and item identifiers. 
For each user $u \in \mathcal{U}$, the model aims to generate the target item identifier(s) given:
\begin{enumerate*}[label=(\arabic*), leftmargin=*]
    \item an interaction history $H_u = {(i_1, b_1), (i_2, b_2), \ldots, (i_N, b_N)}$,
where $i_n \in \mathcal{I}$ and $b_n \in \{\text{src}, \text{rec}\}$ denote the $n$-th item in the interaction history and its interaction type (i.e., search or recommendation), respectively; and
    \item a query $q$, which is provided in the search setting but left empty in recommendation.
\end{enumerate*}


\noindent\textbf{Generative model for unifying S\&R.}
To unify search and recommendation (S\&R) within a generative framework, we formulate both tasks as conditional text generation.
All input information is converted into a unified structured prompt $x$, which is tokenized and fed into the model for instruction tuning.
The model then generates the target item identifier to reflect user preferences through constrained generation.
The training objective minimizes the negative log-likelihood of the user’s preference (i.e., the target item $i$) conditioned on the input instruction $x$ in an autoregressive manner:
\begin{equation}
\label{eqn:gen_loss}
 \mathop{\min}_{\Theta} \{\mathcal{L}({\Theta})=-\sum_{t=1}^{|i|} \log P_{\Theta}(i_t|i_{<t},x)\},
\end{equation}
where $\Theta$ denotes the model parameters, $i_t$ is the $t$-th token of the target item $i$, and $i_{<t}$ represents all tokens preceding $i_t$.
During optimization, we compute gradients and update the model parameters using the Adam optimizer~\cite{kingma2014adam}.
During inference, the model employs beam search to generate the top-$K$ ranked items, which serve as the final S\&R results.
\section{Method}
\label{sec:method}
\begin{figure*}
\setlength{\abovecaptionskip}{0cm}
\setlength{\belowcaptionskip}{0cm}
    \centering    \includegraphics[width=0.75\linewidth]{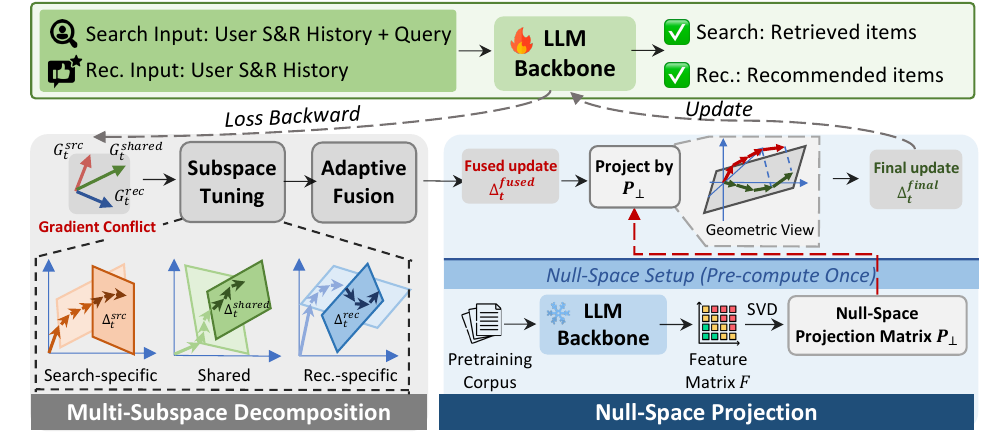}
    \caption{An overview of GEMS. During unified S\&R training, task gradients are routed through multi-subspace decomposition and adaptively fused, then projected onto the null-space of pre-trained knowledge to preserve general-domain understanding.}
\label{fig:method}
\end{figure*}

In this section, we first introduce the core concept of our proposed Gradient Multi-Subspace Tuning (GEMS) framework and its underlying principle of subspace tuning, which enables efficient full-parameter optimization for LLMs (\ref{sec:subspace}). 
We then present the two main components of GEMS, as illustrated in Figure~\ref{fig:method}: multi-subspace decomposition (\ref{sec:multi}) and null-space projection (\ref{sec:null}). 
Finally, we describe the overall training algorithm and analyze the memory and optimization efficiency of GEMS in comparison with the widely used PEFT approach, \textit{i.e.}, LoRA~\citep{hu2022lora} (\ref{sec:training}).

\subsection{Subspace tuning}
\label{sec:subspace}
GEMS is built upon the concept of subspace tuning~\cite{zhao2024galore}, which enables gradient updates within low-rank subspaces, thereby ensuring efficient training in the LLM setting.
At each training step $t$, 
for every trainable layer of the LLM with weight matrix $W_t \in \mathbb{R}^{m \times n}$,
we compute the backpropagated gradient matrix 
$G_t = -\nabla_{W_t}\mathcal{L}(\Theta_t) \in \mathbb{R}^{m \times n}$.
Instead of directly updating $W_t$ using $G_t$ following the Adam optimizer~\cite{kingma2014adam}, as in conventional approaches, we project $G_t$ onto a low-rank subspace spanned by a small set of principal directions of the gradient statistics.
Specifically, we perform singular value decomposition (SVD) on $G_t$:
\begin{equation}
\label{eq:svd}
    G_t = U \Sigma V^\top,
\end{equation}
and retain the top-$r$ singular vectors $U_r \in \mathbb{R}^{m \times r}$ as the basis of the subspace, where $r$ denotes the subspace rank.
The projected gradient is then obtained as:
\begin{equation}
\small
    G_t^{(r)} = U_r^\top G_t.
\end{equation}

Next, optimization is performed entirely within this $r$-dimensional subspace. 
Following the procedure of the Adam optimizer~\cite{kingma2014adam}, we maintain first- and second-order moment estimates $M^{(r)}$ and $V_t^{(r)}$ in this low-rank subspace:
\begin{align}
    Mt^{(r)} &= \beta_1 M_{t-1}^{(r)} + (1 - \beta_1) G_t^{(r)}, \\
    V_t^{(r)} &= \beta_2 V_{t-1}^{(r)} + (1 - \beta_2) \big(G_t^{(r)} \odot G_t^{(r)}\big),
\end{align}
with bias correction:
\begin{equation}
\small
    \hat{M}_t^{(r)} = \frac{M_t^{(r)}}{1 - \beta_1^t}, 
    \quad 
    \hat{V}_t^{(r)} = \frac{V_t^{(r)}}{1 - \beta_2^t},
\end{equation}
where $\beta_1, \beta_2 \in [0,1)$ are decay rates in Adam Optimizer.
The update in the subspace is given by:
\begin{equation}
\small
    \Delta_t^{(r)} = - \eta \cdot \frac{\hat{M}_t^{(r)}}{\sqrt{\hat{V}_t^{(r)}} + \epsilon}.
\end{equation}

Finally, we map this update back into the full parameter space through the basis $U_r$:
\begin{equation}
\label{eq:proj_back}
\small
    \Delta_t = \alpha \cdot U_r \Delta_t^{(r)},
\end{equation}
and apply it to all parameters:
\begin{equation}
    W_{t+1} =W_t + \Delta_t.
\end{equation}
where the scaling factor $\alpha$ controls the magnitude of the low-rank update. 
In this way, optimization proceeds along a few principal directions while still influencing the entire parameter space. 
To further reduce computational cost practically, the SVD is refreshed only once every $T_{\text{svd}}$ steps.

This subspace-based procedure significantly reduces the number of gradient directions, thereby lowering both computational and memory costs. 
Moreover, by filtering gradients through principal directions, subspace tuning suppresses noisy components and enhances training stability, while retaining much of the expressive capacity of full fine-tuning.

\subsection{Multi-subspace decomposition}
\label{sec:multi}
To address gradient conflicts that arise when jointly optimizing search and recommendation objectives, we design a \emph{multi-subspace decomposition} strategy. The key idea is to decompose the optimization space into multiple low-rank subspaces that separate shared and task-specific gradient signals. Directions agreed upon by both tasks are preserved in a shared subspace to promote common representation learning, while task-divergent directions are confined to their respective task-specific subspaces. This separation prevents incompatible updates from competing within the same representational space and thus mitigates destructive interference. Moreover, since each task-specific subspace is derived from distinct gradient statistics, their geometric overlap remains minimal, allowing stable and independent optimization dynamics across tasks. As a result, the model learns generalizable shared representations without compromising task-specific adaptation.

Formally, let $\mathcal{L}_{\text{src}}$ and $\mathcal{L}_{\text{rec}}$ denote the search and recommendation losses, respectively. 
At each training step $t$, we compute the corresponding gradients matrix:
\begin{equation}
\small
\label{eq:gt-specific}
    G_t^{\text{src}} = -\nabla_{W_t} \mathcal{L}_{\text{src}}(\Theta), 
    \quad
    G_t^{\text{rec}} = -\nabla_{W_t} \mathcal{L}_{\text{rec}}(\Theta).
\end{equation}
We further define the shared gradient as:
\begin{equation}
\label{eq:gt-shared}
    G_t^{\text{shared}} = -\nabla_{W_t} (\mathcal{L}_{\text{src}}(\Theta)+\mathcal{L}_{\text{rec}}(\Theta)).
\end{equation}

\noindent\textbf{Shared and task-specific subspaces.} 
Based on these gradients, GEMS constructs one shared subspace and two task-specific subspaces to disentangle common and distinct optimization directions. 
The \textbf{shared subspace} captures gradient directions beneficial to both tasks, including general user preference patterns (e.g., broad interest categories) and semantic information from interacted items (e.g., textual representations). 
The \textbf{search-specific subspace} focuses on signals unique to search, such as modeling semantic intent in queries and aligning user queries with item content~\cite{wu2024generative}. 
The \textbf{recommendation-specific subspace} emphasizes signals unique to recommendation, such as user long-term preferences and collaborative filtering patterns from historical interactions~\cite{zhao2025model,lin2026bringing}. 
Following the subspace tuning procedure in Section~\ref{sec:subspace}, the corresponding updates $\Delta_t^{\text{shared}}$, $\Delta_t^{\text{src}}$, and $\Delta_t^{\text{rec}}$ are derived from their respective gradients $G_t^{\text{shared}}$, $G_t^{\text{src}}$, and $G_t^{\text{rec}}$ according to Eq.~\ref{eq:svd}--\ref{eq:proj_back}.

\noindent\textbf{Adaptive fusion of subspaces.}
After obtaining the gradient updates from each subspace, 
the three projected gradients are then fused through an adaptive gating mechanism.  
In this design, the shared subspace always contributes to the final update with a fixed weight of $1$, ensuring that task-invariant signals are consistently preserved.
By contrast, the search- and recommendation-specific subspaces are combined with dynamic weights $\alpha_{\text{src}}, \alpha_{\text{rec}} \geq 0$ that satisfy
\begin{equation}
    \alpha_{\text{src}} + \alpha_{\text{rec}} = 1.
\end{equation}
The fused update applied to the parameter set is therefore
\begin{equation}
\label{eq:fuse}
    \Delta_t^{\text{fused}} = \Delta_t^{\text{shared}}
    + \alpha_{\text{src}} \, \Delta_t^{\text{src}}
    + \alpha_{\text{rec}} \, \Delta_t^{\text{rec}}.
\end{equation}
and applied to all parameters:
\begin{equation}
    W_{t+1} =W_t + \Delta_t^{\text{fused}}.
\end{equation}

To determine the gating weights, we design a lightweight neural network conditioned on task-level statistics that reflect the relative learning states of the two tasks.  
In particular, we extract three types of normalized ratios: 
(i) the relative magnitudes of task losses, 
(ii) the relative gradient norms of the two tasks, and 
(iii) the relative sample sizes within the current batch.  
These ratios collectively form a feature vector 
\begin{equation}
    z = \left[s^{\text{loss}}, \; s^{\text{grad}}, \; s^{\text{sample}} \right] .
\end{equation}

The adaptive gating network $f_{\phi}(\cdot)$ is implemented as a two-layer perceptron with ReLU activations:
\begin{equation}
\small
    h = \sigma\!\left(W_1 z + b_1\right), \quad
    o = W_2 h + b_2,
\end{equation}
where $\sigma(\cdot)$ is the ReLU function, and $\phi = \{W_1, b_1, W_2, b_2\}$ are learnable parameters.
This design enables the gating module to nonlinearly model the relationships among multiple task indicators and to infer which subspace should receive greater emphasis at each step.
In essence, the gating network serves as a meta-controller that monitors the training dynamics of both tasks and adaptively allocates learning capacity: when one task exhibits higher loss or unstable gradients, its corresponding subspace receives a smaller weight, thereby stabilizing joint optimization and preventing one task from dominating the shared representation learning.

Finally, the gating weights are obtained by a temperature-scaled softmax:
\begin{equation}
    \alpha = \text{softmax}\!\left(\frac{o}{\tau}\right),
\end{equation}
where $\tau$ is a gate temperature factor and controls the sharpness of the gating distribution.
Thus, $\alpha = [ \alpha_{\text{src}}, \alpha_{\text{rec}}]$ defines the adaptive combination coefficients for the three subspace gradients in Eq.~(\ref{eq:fuse}).
This enables the framework to adaptively balance shared and task-specific updates during training, ensuring stable convergence and effective coordination between S\&R learning.

\subsection{Null-space projection}
\label{sec:null}
As stated in the Section~\ref{sec:introduction}, PEFT methods may cause a shift in the user intent understanding, distribution of pre-trained knowledge, leading to general-domain knowledge disruption and erosion of the semantic and reasoning abilities of the backbone LLM.  
To mitigate this problem, we introduce a \textit{null-space projection} that preserves general-domain knowledge while still enabling effective adaptation to unified S\&R tasks.  

The key idea is to identify the principal representation space of the backbone LLM, which corresponds to its general-domain knowledge, and to prevent fine-tuning updates from moving along these dominant directions.
To achieve this, updates are projected into the complementary subspace (the null space), which poses a lower risk of disturbing pre-trained semantics~\cite{fang2024alphaedit}.

To estimate the principal directions of the backbone LLM's pre-trained representation space, we construct a feature matrix $F \in \mathbb{R}^{n \times C}$ for each fine-tuned layer by feeding a representative general-domain corpus $\mathcal{C}$ into the frozen backbone and stacking the resulting layer input hidden states as columns of $F$.
Here, $n$ denotes the dimensionality of the layer’s hidden representation (the input dimension of the corresponding weight matrix $W_t \in \mathbb{R}^{m \times n}$), and $C$ is the number of input instances from $\mathcal{C}$.
We compute the covariance matrix $F F^\top$ and perform SVD:
\begin{equation}
   F F^\top = U_{\text{pre}} \Sigma_{\text{pre}} U_{\text{pre}}^\top,
\end{equation}
where columns of $U_{\text{pre}}$ are orthonormal singular vectors ordered by descending singular values.
Intuitively, the top-$k$ singular vectors $U_{\text{pre}}^k \in \mathbb{R}^{n \times k}$ capture the dominant semantic directions most critical to general reasoning and semantic understanding.

Following this, we define the projection matrix onto the null-space of $U_{\text{pre}}^k$
(the top-$k$ singular vectors of $U_{\text{pre}}$) as:
\begin{equation}
    P_{\perp} = I - U_{\text{pre}}^k U_{\text{pre}}^{k\top},
\end{equation}
which removes any component of an update that aligns with the dominant pre-trained directions. 
Since the projection matrix $P_{\perp}$ depends solely on the general-domain knowledge,
it only needs to be computed once per layer during the preparation stage before training.
Given a fused gradient update $\Delta_t^{\text{fused}}$ at training step $t$, the final update is obtained by:
\begin{equation}
    \Delta_t^{\text{final}} = \Delta_t^{\text{fused}} P_{\perp}.
\end{equation}

\noindent This procedure ensures that gradient updates avoid interfering with the principal directions associated with pre-trained representations, thereby reducing knowledge shift and preserving the semantic structures acquired during pre-training.
As a result, the model adapts to S\&R tasks mainly along directions less aligned with prior knowledge, striking a balance between efficient fine-tuning and knowledge preservation.


\begin{algorithm}[t]
\caption{Training algorithm of GEMS}
\label{alg:train}
\begin{algorithmic}[1]

\Require LLM Layer weight $W\!\in\!\mathbb{R}^{m\times n}$ ($m\le n$). Step size $\eta$, scale factor $\alpha$, rank $r$, SVD refresh step $T_{\mathrm{svd}}$, gate temperature factor $\tau$, pretrained basis $\mathbf{U}_{\text{ref}}^k$.
\State Initialize moments $M^{(\mathrm{src})},M^{(\mathrm{rec})},M^{(\mathrm{shared})}\in\mathbb{R}^{r\times n}\gets \mathbf{0}$, $V^{(\mathrm{src})},V^{(\mathrm{rec})},V^{(\mathrm{shared})}\in\mathbb{R}^{r\times n}\gets \mathbf{0}$
\State Initialize global step counter $t \gets 0$


\Function{SubspaceTune}{type, $\mathbf{g}, t$}
  \State \Comment{Look up and maintain per-type states: $U^{(\text{type})}_r$, $M^{(\text{type})}$, $V^{(\text{type})}$}
  \State \textbf{if} $t \bmod T_{\mathrm{svd}}=0$ \textbf{ then} refresh $U^{(\text{type})}_r$ via SVD as in Eq.(\ref{eq:svd})
  \State Compute $\Delta_t^{(\text{type})}$ according to Eq.(\ref{eq:svd})-(\ref{eq:proj_back}) given $\mathbf{g}$
  \State \Return $\Delta_t^{(\text{type})}$
\EndFunction

\Function{NullProject}{$\Delta,\,\mathbf{U}_k$}
  \State \Return $\Delta(I - \mathbf{U}_k\mathbf{U}_k^{\top})\,$ \Comment{null-space projection}
\EndFunction

\Repeat
  \State Sample mini-batch $\{(u,H_u,q,i^{*})\}_{b=1}^B$ \Comment{$q\neq\varnothing$: search;\; $q=\varnothing$: recommendation}
\State Construct model input $x$ from $(H_u, q)$ and compute $\mathcal{L}_{\mathrm{src}},\,\mathcal{L}_{\mathrm{rec}}$ according to Eq.(\ref{eqn:gen_loss})
  \State 
  Compute $\mathbf{G}_t^{\mathrm{src}}$, $\mathbf{G}_t^{\mathrm{rec}}$, $\mathbf{G}_t^{\mathrm{shared}}$ according to Eq.(\ref{eq:gt-specific})-(\ref{eq:gt-shared})

  \State $\Delta_t^{\mathrm{src}}\!\gets\!\Call{SubspaceTune}{\mathrm{src},\,\mathbf{G}_t^{\mathrm{src}},\,t}$
  \State $\Delta_t^{\mathrm{rec}}\!\gets\!\Call{SubspaceTune}{\mathrm{rec},\,\mathbf{G}_t^{\mathrm{rec}},\,t}$
  \State $\Delta_t^{\mathrm{shared}}\!\gets\!\Call{SubspaceTune}{\mathrm{shared},\,\mathbf{G}_t^{\mathrm{shared}},\,t}$

  \State $\Delta_t^{\mathrm{fuse}}\!\gets\!\Delta_t^{\mathrm{shared}}+\alpha_{\mathrm{src}}\Delta_t^{\mathrm{src}}+\alpha_{\mathrm{rec}}\Delta_t^{\mathrm{rec}}$ \Comment{adaptive fusion}

  \State $\Delta_t^{\mathrm{final}}\!\gets\!\Call{NullProject}{\Delta_t^{\mathrm{fuse}},\,\mathbf{U}_{\text{ref}}^k}$ \Comment{null-space projection}
  \State $W\!\gets\!W+\Delta_t^{\mathrm{final}}$; $t\!\gets\!t+1$ \Comment{parameter update}
\Until{convergence}
\State \Return $W$
\end{algorithmic}
\end{algorithm}
\subsection{Training algorithm}
\label{sec:training}
We list the detailed training algorithm in Algorithm~\ref{alg:train}.
At each step $t$, a mini-batch is sampled to compute task-specific losses $\mathcal{L}{\text{src}}$ and $\mathcal{L}{\text{rec}}$, yielding gradient matrices $\boldsymbol{G}_t^{\text{src}}$ and $\boldsymbol{G}_t^{\text{rec}}$, as well as a shared gradient $\boldsymbol{G}t^{\text{shared}}$. 
These gradients define three low-rank subspaces (i.e., shared, search-specific, and recommendation-specific), whose bases are obtained via SVD and refreshed every $T{\text{svd}}$ steps.
Parameter updates ($\Delta_t^{\text{shared}}$, $\Delta_t^{\text{src}}$, $\Delta_t^{\text{rec}}$) are computed by projecting gradients into these subspaces and are optimized using independent Adam optimizers.
An adaptive gating mechanism then fuses their contributions into $\Delta_t^{\text{fused}}$, which undergoes a null-space projection to obtain $\Delta_t^{\text{final}}$, ensuring orthogonality to the general-domain knowledge space.
The final update $\Delta_t^{\text{final}}$ is applied to the model parameters $W_t$, and the process repeats until convergence.

\section{Experiments}
\label{sec:exp}We address the following research questions:
\begin{enumerate*}[label=(\textbf{RQ\arabic*}),leftmargin=*,nosep]
    \item Does GEMS outperform state-of-the-art baselines on the S\&R tasks? 
    \item Do the components of GEMS contribute to its effectiveness?
    \item Can GEMS mitigate gradient conflicts and shifts in user intent understanding for S\&R?
    \item How do hyperparameters influence the performance of GEMS?
\end{enumerate*}

\subsection{Experimental setup}
\label{sec:setting}

\label{subsec:data{statkstics}}

\begin{table}[t]
\small
\setlength{\abovecaptionskip}{0cm}
\setlength{\belowcaptionskip}{0cm}
\centering
\caption{Statistics of two datasets. \#Inter-S and \#Inter-R denote the interaction number of search behavior and recommendation behavior, respectively.}
\label{tab:datasets}
\resizebox{\linewidth}{!}{%
\begin{tabular}{ccccccc}
\toprule
\textbf{Dataset}     & \textbf{\#User} & \textbf{\#Item} & \textbf{\#Inter-S} & \textbf{\#Inter-R} & \textbf{Density-S} & \textbf{Density-R} \\ \midrule 
\textbf{Qilin}    & 3,816        & 275,515             & 16,626                       & 101,085    & 0.00158\% & 0.00962\%     \\
\textbf{Amazon} & 62,909         & 158,296        & 389,342                         & 451,301     & 0.00391\%  & 0.00453\%     \\
\bottomrule 
\end{tabular}}
\end{table}

\noindent\textbf{Datasets.} 
We evaluate GEMS on two publicly available datasets containing both S\&R histories: 
\begin{enumerate*}
    \item \textbf{Qilin}\footnote{\url{https://github.com/RED-Search/Qilin.}}: It is a dataset collected from a social media application, a popular content-sharing platform that integrates both S\&R behaviors, including real queries and item descriptions. 
    \item \textbf{Amazon}\footnote{\url{https://github.com/QingyaoAi/Amazon-Product-Search-Datasets}.}: Following previous work~\cite{si2023search,shi2024unisar,ai2017learning,shi2025unified}, we use the semi-synthetic dataset based on Amazon product review data. Specifically, we use the \emph{Electronics} 5-core subset, which provides reviews and ratings for electronic devices. 
\end{enumerate*}
The dataset statistics are shown in Table~\ref{tab:datasets}.
Details of data processing and splitting are provided in our code repository.
To the best of our knowledge, these are the only publicly available unified S\&R datasets that provide both rich item semantics and explicit natural-language search queries, which are indispensable for leveraging the semantic reasoning capabilities of LLMs in our setting.

\noindent\textbf{Baselines.}
We compare GEMS with representative non-generative baselines and generative methods for recommendation, search, and unified S\&R, together with mainstream PEFT approaches.
\noindent
\textbf{Recommendation baselines:}  
(1) \textbf{NCF}~\cite{He2017Neural} replaces the inner product in matrix factorization with a neural architecture to model complex user--item interactions.  (2) \textbf{TIGER}~\cite{rajput2023recommender} assigns discrete Semantic IDs to items and trains a seq2seq model to predict the next item’s ID in a user’s history.  
(3) \textbf{LETTER}~\cite{wang2024learnable} integrates hierarchical semantics, collaborative signals, and code assignment diversity into item identifiers for LLM-based generative recommendation.  
\noindent
\textbf{Search baselines:}  
(4) \textbf{ANCE}~\cite{xiong2020approximate} improves dense retrieval by mining hard negatives via an asynchronously updated ANN index.  
(5) \textbf{WebUltron}~\cite{zhou2023webultron} introduces an end-to-end generative retrieval framework with semantically rich item identifiers.  
(6) \textbf{GenRet}~\cite{sun2023learning} tokenizes items into learnable semantic identifiers.  
\noindent
\textbf{Unifying baselines:}  
(7) \textbf{UnifiedSSR}~\cite{xie2024unifiedssr} shares information across scenarios/views and models user intent with self-supervised session discovery.  
(8) \textbf{BSR}~\cite{penha2024bridging} uses atomic item IDs to jointly train generative models for search and recommendation.  
(9) \textbf{Sem-BSR}~\cite{penha2025semantic} learns effective Semantic IDs by jointly fine-tuning a bi-encoder on both tasks.  
(10) \textbf{GenSAR}~\cite{shi2025unified} designs dual-purpose semantic and collaborative item identifiers to unify generative search and recommendation.  

\noindent
\textbf{PEFT methods:}  
(11) \textbf{LoRA}~\cite{hu2022lora} reparameterizes weight matrices into low-rank components and trains lightweight adapters, enabling efficient adaptation without modifying the entire parameter space.
(12) \textbf{LoRA-MoE}~\cite{dou2023loramoe} introduces a plugin Mixture-of-Experts (MoE) architecture with LoRA experts and localized balancing constraints to prevent LLMs' world knowledge forgetting during fine-tuning.

\noindent
\textbf{Backbone baseline:}
(13) \textbf{Vanilla backbone}: 
In this baseline, the backbone model is directly applied to S\&R without any task-specific fine-tuning. It serves as a lower bound that quantifies the performance achievable by the pretrained model alone.

\noindent\textbf{Implementation details.} 
Following prior work~\cite{shi2024unisar,xie2024unifiedssr,yao2021user}, we randomly sample 99 negative items per user and combine them with the ground-truth item to form candidate lists. All models rank these candidates, and we evaluate top-$K$ S\&R performance using Hit@$K$ and NDCG@$K$ with $K=\{5,10\}$.  
Hyperparameters for baselines are tuned within the ranges reported in their original papers. 
For fairness, we apply constrained beam search to all generative methods, ensuring outputs are restricted to the candidate set. 
We evaluate generative methods (including our model, generative baselines, and the vanilla backbone) under two backbone LLMs: Flan-T5-base\footnote{\url{https://huggingface.co/google/flan-t5-base}.} and Qwen2.5-3B-Instruct\footnote{\url{https://huggingface.co/Qwen/Qwen2.5-3B-Instruct}.}.
For baselines under Flan-T5-base, we use full fine-tuning for all generative methods. 
For baselines under Qwen2.5-3B-Instruct, we adopt parameter-efficient training with LoRA to reduce cost.
We further include LoRA-MoE~\cite{dou2023loramoe} as a stronger PEFT baseline for Sem-BSR-MoE to test whether gains hold under more expressive PEFT. Non-generative baselines (e.g., NCF and ANCE) do not use LLM backbones and follow standard training and evaluation.
For null-space projection, we use the Wikipedia dataset\footnote{\url{https://huggingface.co/datasets/wikimedia/wikipedia/viewer/20231101.en.}} as the general-domain corpus $\mathcal{C}$, uniformly sampling $C=100{,}000$ documents to estimate the pretraining subspace. 
We set SVD refresh step $T_\text{svd}$ to 200.
We set $k$, the number of retained principal directions, equal to the subspace rank to match projection capacity with the tunable update space and avoid extra hyperparameters.
The best hyperparameters are selected through grid search over the following ranges:
the scale factor $\alpha$ is tuned in $\{0.5, 1, 2, 3, 4\}$, the gate temperature factor $\tau$ in $\{0.1, 0.5, 1, 2, 3\}$, and the subspace rank\footnote{Each task-specific subspace rank is set to half of the shared subspace rank. 
This design encourages compact specialization by allowing task-specific parameters to capture residual variations around the shared representation while preventing over-parameterization.} in $\{256, 512, 1024\}$.

\begin{table*}[t]
\caption{Results on S\&R. For all generative methods (including our model and generative baselines), the backbone model is T5-base. The best and second-best results are highlighted in bold and underlined fonts, respectively. * indicates that the best result is statistically significantly better than the second-best (t-test, $p$ < 0.01).}
\label{tab:t5_result}
\centering
\setlength{\tabcolsep}{5pt}
\renewcommand{\arraystretch}{1.05}
\small
\begin{tabular}{llcccccccc}
\toprule
& & \multicolumn{4}{c}{\textbf{Qilin}} & \multicolumn{4}{c}{\textbf{Amazon}} \\
\cmidrule(lr){3-6} \cmidrule(lr){7-10}
\textbf{Task} & \textbf{Method} & \textbf{Hit@5} & \textbf{Hit@10} & \textbf{NDCG@5} & \textbf{NDCG@10}
              & \textbf{Hit@5} & \textbf{Hit@10} & \textbf{NDCG@5} & \textbf{NDCG@10} \\
\midrule
\multirow{9}{*}{\textbf{Recommendation}}
 & Vanilla Backbone & 0.0997 & 0.1475 & 0.0619 & 0.0851 & 0.0546 & 0.0974 & 0.0312 & 0.0459 \\
 & NCF        & 0.1497 & 0.2019 & 0.1267 & 0.1433  & 0.1718 & 0.2035 & 0.1009 & 0.1273 \\
 & TIGER      & \underline{0.2548} & \underline{0.3052} & \underline{0.1971} & \underline{0.2091} & \underline{0.2019} & 0.2581 & \underline{0.1494} & \underline{0.1675} \\
 & LETTER     & 0.2116 & 0.2577 & 0.1605 & 0.1723 & 0.2014 & 0.2582 & 0.1491 & 0.1674 \\
 & UnifiedSSR & 0.1531 & 0.2171 & 0.1325 & 0.1446 & 0.1477 & 0.1986 & 0.0923 & 0.1313 \\
 & BSR        & 0.2089 & 0.2432 & 0.1587 & 0.1673 & 0.2004 & 0.2568 & 0.1474 & 0.1656 \\
 & Sem-BSR    & 0.2437 & 0.2936 & 0.1904 & 0.2033 & \underline{0.2019} & \underline{0.2584} & 0.1484 & 0.1664 \\
 & GenSAR     & 0.2168 & 0.2527 & 0.1678 & 0.1771 & 0.2008 & 0.2575 & 0.1478 & 0.1661 \\
 & \textbf{Ours} & \textbf{0.4285*} & \textbf{0.5121*} & \textbf{0.3251*} & \textbf{0.3465*}
                 & \textbf{0.4025*} & \textbf{0.5159*} & \textbf{0.2975*} & \textbf{0.3341*} \\
\midrule
\multirow{9}{*}{\textbf{Search}}
 & Vanilla Backbone & 0.0348 & 0.0471 & 0.0208 & 0.0312 & 0.0572 & 0.1053 & 0.0319 & 0.0567 \\
 & ANCE        & 0.0209 & 0.0324 & 0.0114 & 0.0153 & 0.0487 & 0.0979 & 0.0285 & 0.0441 \\
 & WebUltron   & 0.0228 & \underline{0.0390} & 0.0136 & 0.0185 & 0.2213 & 0.2554 & 0.1962 & 0.2071 \\
 & GenRet      & \underline{0.0262} & 0.0378 & \underline{0.0156} & 0.0194 & \underline{0.4198} & 0.4205 & \underline{0.3995} & \underline{0.3998} \\
 & UnifiedSSR  & 0.0138 & 0.0251 & 0.0097 & 0.0227 & 0.1866 & 0.2481 & 0.1305 & 0.1558 \\
 & BSR         & 0.0153 & 0.0290 & 0.0137 & 0.0259 & 0.2058 & \underline{0.3215} & 0.1702 & 0.2075 \\
 & Sem-BSR     & 0.0261 & 0.0606 & 0.0152 & \underline{0.0263} & 0.2023 & 0.2603 & 0.1496 & 0.1683 \\
 & GenSAR      & 0.0141 & 0.0286 & 0.0090 & 0.0136 & 0.4032 & \underline{0.4608} & 0.3170 & 0.3686 \\
 & \textbf{Ours} & \textbf{0.1511*} & \textbf{0.1922*} & \textbf{0.0989*} & \textbf{0.1122*}
                 & \textbf{0.8399*} & \textbf{0.8456*} & \textbf{0.8000*} & \textbf{0.8018*} \\
\bottomrule
\end{tabular}
\end{table*}

\begin{table*}[t]
\caption{Results on S\&R. Backbone model is Qwen-3B. The best and second-best results are highlighted in bold and underlined fonts, respectively. * indicates that the best result is statistically significantly better than the second-best (t-test, $p$ < 0.01).}
\label{tab:qwen_result}
\centering
\setlength{\tabcolsep}{5pt}
\renewcommand{\arraystretch}{1.05}
\small
\begin{tabular}{llcccccccc}
\toprule
& & \multicolumn{4}{c}{\textbf{Qilin}} & \multicolumn{4}{c}{\textbf{Amazon}} \\
\cmidrule(lr){3-6} \cmidrule(lr){7-10}
\textbf{Task} & \textbf{Method} & \textbf{Hit@5} & \textbf{Hit@10} & \textbf{NDCG@5} & \textbf{NDCG@10}
              & \textbf{Hit@5} & \textbf{Hit@10} & \textbf{NDCG@5} & \textbf{NDCG@10} \\
\midrule
\multirow{8}{*}{\textbf{Recommendation}}
 & Vanilla Backbone & 0.0712 & 0.0985 & 0.0512 & 0.0603 & 0.0624 & 0.0856 & 0.0415 & 0.0498 \\
 & TIGER        & 0.2469 & 0.3008 & 0.1858 & 0.2001 & \underline{0.1886} & 0.2487 & 0.1375 & 0.1571 \\
 & LETTER       & 0.1845 & 0.2446 & 0.1338 & 0.1532 & 0.1845 & 0.2446 & 0.1338 & 0.1532 \\
 & BSR          & 0.1087 & 0.1356 & 0.0813 & 0.0881 & 0.0837 & 0.1215 & 0.0598 & 0.0703 \\
 & Sem-BSR      & 0.2432 & 0.2891 & 0.1893 & 0.2009 & 0.1300 & 0.1938 & 0.0873 & 0.1078 \\
 & GenSAR       & 0.1168 & 0.1625 & 0.0832 & 0.0971 & 0.0916 & 0.1382 & 0.0625 & 0.0775 \\
 & Sem-BSR-MoE  & \underline{0.3061} & \underline{0.3419} & \underline{0.2491} & \underline{0.2562} & 0.1812 & \underline{0.2799} & \underline{0.1806} & \underline{0.1823} \\
 & \textbf{Ours} & \textbf{0.3499*} & \textbf{0.4219*} & \textbf{0.2648*} & \textbf{0.2837*}
                 & \textbf{0.3686*} & \textbf{0.4743*} & \textbf{0.2717*} & \textbf{0.3056*} \\
\midrule
\multirow{8}{*}{\textbf{Search}}
 & Vanilla Backbone & 0.0093 & 0.0142 & 0.0056 & 0.0078 & 0.1152 & 0.1589 & 0.0812 & 0.0954 \\
 & WebUltron    & 0.0369 & 0.0535 & 0.0218 & 0.0271 & 0.1754 & 0.2423 & 0.1243 & 0.1459 \\
 & GenRet       & \underline{0.0510} & \underline{0.0776} & \underline{0.0306} & \underline{0.0393} & 0.1598 & 0.2309 & 0.1107 & 0.1336 \\
 & BSR          & 0.0124 & 0.0211 & 0.0075 & 0.0102 & 0.1732 & 0.2391 & 0.1237 & 0.1445 \\
 & Sem-BSR      & 0.0279 & 0.0469 & 0.0196 & 0.0255 & 0.3132 & 0.3515 & \underline{0.2719} & 0.2843 \\
 & GenSAR       & 0.0155 & 0.0224 & 0.0101 & 0.0123 & 0.1997 & 0.2259 & 0.1785 & 0.1868 \\
 & Sem-BSR-MoE  & 0.0328 & 0.0532 & 0.0258 & 0.0313 & \underline{0.3291} & \underline{0.3729} & \textbf{0.2802} & \underline{0.2893} \\
 & \textbf{Ours} & \textbf{0.0552*} & \textbf{0.0780*} & \textbf{0.0374*} & \textbf{0.0447*}
                 & \textbf{0.3510*} & \textbf{0.4570*} & 0.2596 & \textbf{0.2938*} \\
\bottomrule
\end{tabular}
\end{table*}

\subsection{Results on S\&R (RQ1)}
\label{sec:overall_performance}
We first conduct a comprehensive comparison under our default setting with Flan-T5-base as the backbone (Table~\ref{tab:t5_result}), where we compare GEMS against a broad set of baselines. 
To further verify backbone robustness, we deploy all generative methods (including our model, generative baselines, and the vanilla backbone) on a second backbone, Qwen2.5-3B-Instruct (Table~\ref{tab:qwen_result}). 
Our experiments yield three key observations:  
\begin{enumerate*}[leftmargin=*]
    \item Unifying S\&R remains challenging. Unified baselines often underperform specialized models, underscoring severe gradient conflicts between the two objectives, where updates that benefit one task may degrade the other when a shared representation is enforced. This effect is particularly pronounced with T5-base, where full fine-tuning exposes all parameters (embeddings, attention, and feed-forward layers) to both tasks, amplifying cross-task interference.  
    \item GEMS achieves the highest performance across nearly all datasets and tasks. The gains are particularly notable under T5-base, where it outperforms the best specialized baselines. This superior performance can be attributed to GEMS's multi-subspace decomposition and null-space projection. 
    \item Compared to mainstream PEFT baselines and PEFT–MoE variants, GEMS mostly achieves the best results. These results demonstrate that GEMS can provide greater performance gains while maintaining efficiency. Notably, its ability to outperform MoE-enhanced PEFT suggests that merely increasing routing capacity is insufficient to resolve cross-task interference.
\end{enumerate*}


\begin{figure}
  \centering
  \subfigure{
    \includegraphics[width=0.48\linewidth]{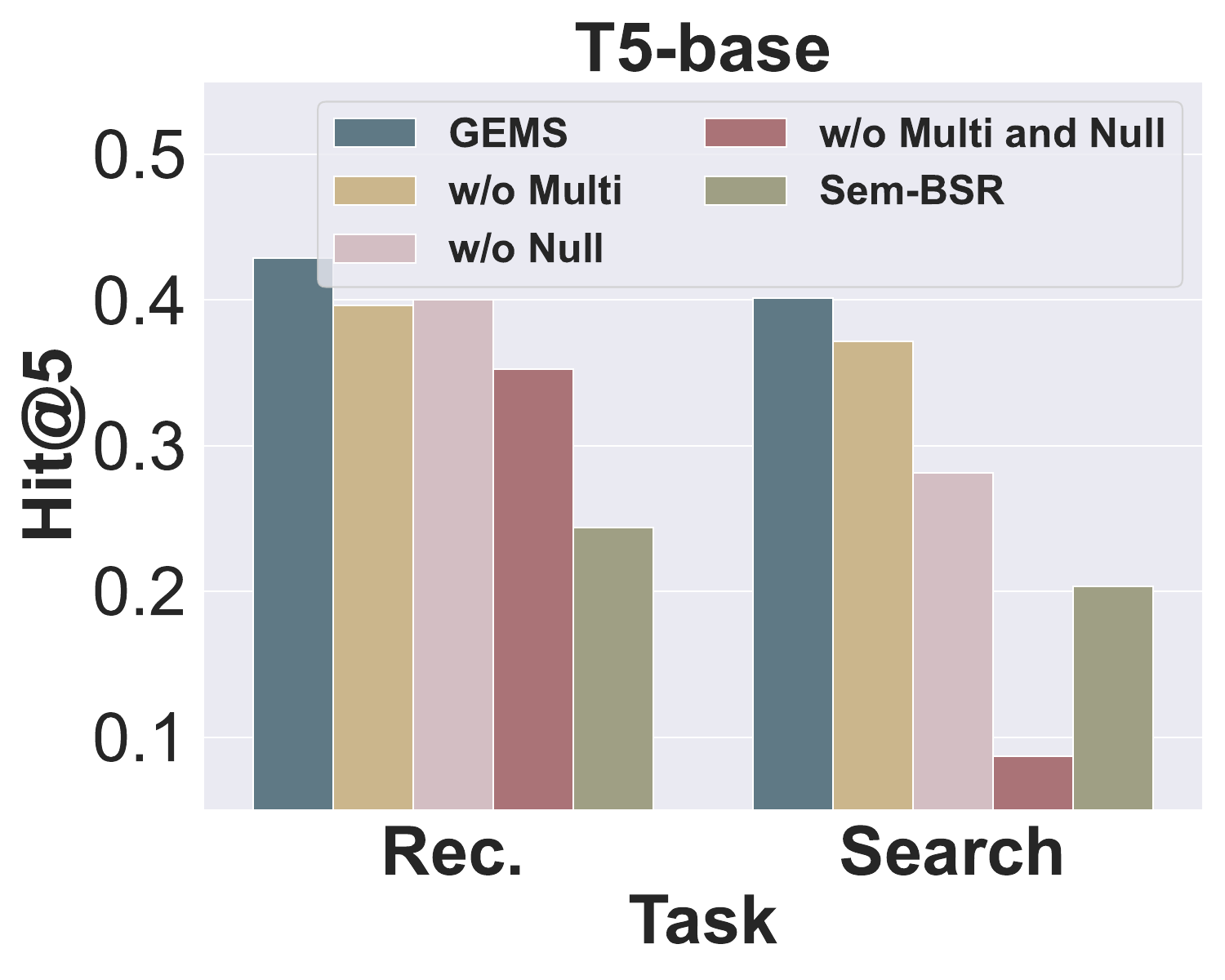}}
  \hfill
  \subfigure{
    \includegraphics[width=0.48\linewidth]{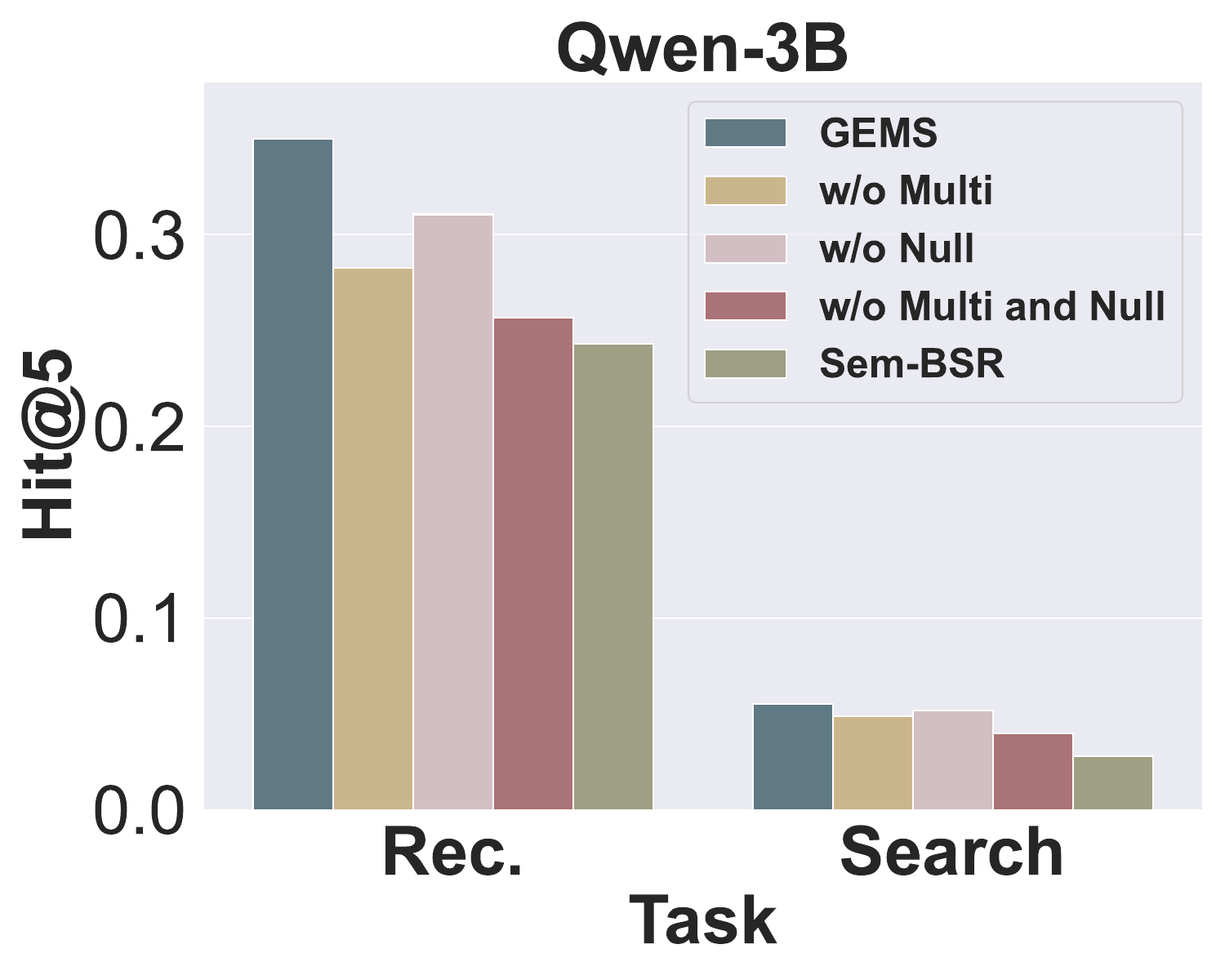}}
  \caption{Ablation study of GEMS on Qilin.}
  \label{fig:abl}
\end{figure}

\begin{figure}[t]  
    \centering    
    \includegraphics[width=\linewidth]{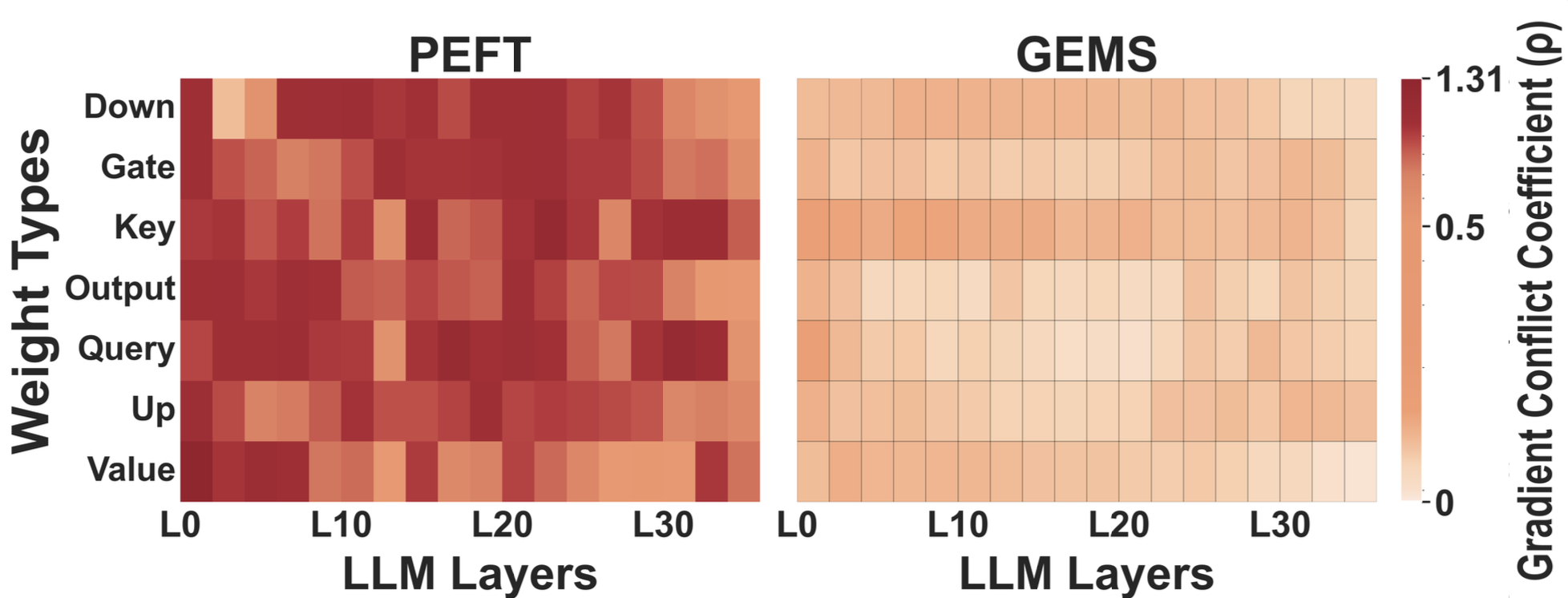}
  \caption{Gradient conflict heatmap analysis of PEFT and GEMS using Qwen-3B on the Qilin dataset.}
    \label{fig:grad}
\end{figure}

\subsection{In-depth Analysis}
\label{sec:in-depth-exp}
After addressing RQ1, we conduct a detailed analysis of GEMS. Specifically, we examine the contributions of multi-subspace decomposition and null-space projection to its performance (\ref{subsubsec:ablation study}), evaluate its ability to mitigate gradient conflicts (\ref{subsubsec:GCA}) and preserve user intent understanding (\ref{subsubsec:UIUPA}), and analyze its sensitivity to hyperparameters (\ref{subsec:hyperparameters}).

\subsubsection{Ablation study (RQ2).} 
\label{subsubsec:ablation study}
To better understand the contribution of each component within GEMS, we conduct an ablation study by progressively removing its two core modules: 
multi-subspace decomposition and null-space projection. 
Figure~\ref{fig:abl} shows the results for both backbone models also compared with strong unifying S\&R baseline Sem-BSR, revealing the following key findings:
(1) \textbf{Effect of multi-subspace decomposition}:
When we disable the multi-subspace decomposition and instead optimize over a single shared subspace, we observe a substantial degradation in performance across both S\&R, which confirms that explicitly disentangling shared and task-specific signals into complementary subspaces is essential for mitigating interference and maintaining stable optimization.
(2) \textbf{Effect of null-space projection}:
In the setting without the null-space projection step, the model achieves moderate improvements over baselines but lags behind the full GEMS,
which highlights the importance of constraining updates to remain orthogonal to pre-trained knowledge representations in order to preserve the LLM's language understanding ability.
(3) \textbf{Effect of subspace tuning.}
We also evaluate the version that applies only single-subspace tuning, removing both modules.
Overall, subspace tuning demonstrates strong performance by optimizing within an effective low-rank subspace.
However, in certain scenarios, gradient conflicts diminish this advantage, leading to cases where Sem-BSR outperforms it, further underscoring the effectiveness of Multi-Subspace Decomposition.

\subsubsection{Gradient conflict analysis (RQ3).} 
\label{subsubsec:GCA}
In this section, we empirically analyze gradient conflicts in GEMS to assess how effectively our method alleviates them. 
We quantify the degree of conflict using the \textbf{Gradient Conflict Coefficient ($\rho$)}, which measures how strongly the task gradients align or oppose each other during training. 
Given the gradients of the search task $g_{\text{src}}$ and the recommendation task $g_{\text{rec}}$,  $\rho$ is defined as:
\begin{equation}
\text{$\rho$} = 1 - \frac{g_{\text{src}} \cdot g_{\text{rec}}}{\|g_{\text{src}}\| \, \|g_{\text{rec}}\|}.
\end{equation}
We compute $\rho$ per layer on each mini-batch (normalized per batch) and report layerwise results by averaging over evaluation samples.
A higher coefficient indicates stronger conflicts, while a lower coefficient reflects better compatibility between the two objectives. 
As illustrated in Figure~\ref{fig:grad} and Figure~\ref{fig:intro}(a), we find that: 
(1) GEMS consistently achieves lower gradient conflict coefficients across all LLM layers than the PEFT baseline (LoRA), indicating that it effectively mitigates competing optimization signals between S\&R and facilitates smoother training dynamics. 
This finding validates the effectiveness of the multi-subspace decomposition mechanism. 
Quantitatively, GEMS alleviates over 85\% (up to 88\%) of average conflict magnitudes across all weight types.
(2) In PEFT, high-conflict regions predominantly occur in the \textit{Query} and \textit{Key} layers, where components that control attention alignment and are particularly sensitive to multi-task interference. 
In contrast, GEMS maintains uniformly low conflict magnitudes across layers, demonstrating more stable optimization behavior.


\begin{table}[t]
\centering
{\small
\caption{User intent understanding preservation analysis under Qwen-3B. We report the percentage of ``correct-before $\rightarrow$ incorrect-after'' (lower is better). $\Delta$ denotes the absolute reduction compared to the BSR baseline.}
\label{tab:intent-preservation}
\begin{tabular}{lcccc}
\toprule
\textbf{Dataset} & \textbf{Task} & \textbf{PEFT} & \textbf{Ours } & \textbf{$\Delta$} \\
\midrule
\multirow{2}{*}{\textbf{Qilin}} 
& Rec.   & 21.1\% & \textbf{10.6\%} & -10.5\% \\
& Search & 24.8\% & \textbf{12.9\%} & -11.9\% \\
\midrule
\multirow{2}{*}{\textbf{Amazon}} 
& Rec.   & 18.9\% & \textbf{7.9\%}  & -11.0\% \\
& Search & 26.2\% & \textbf{11.7\%} & -14.5\% \\
\bottomrule
\end{tabular}
}
\vspace{-3mm}
\end{table}

\subsubsection{User intent understanding preservation analysis (RQ3).} 
\label{subsubsec:UIUPA}
A key concern raised in the introduction is that adapting LLMs with PEFT in unified S\&R can distort user intent understanding by overfitting to finetuned data and undermining the inherent language understanding and reasoning abilities of LLMs.
To investigate whether GEMS better preserves such capability, we conduct an analysis comparing prediction consistency before and after fine-tuning. 
Concretely, we report the proportion of cases in which the base LLMs was correct but became incorrect after tuning.
As shown in Table~\ref{tab:intent-preservation}, PEFT (LoRA) exhibits a substantial share of disrupted cases across datasets, indicating that parameter-efficient updates can overwrite pre-trained representations in undesirable ways. 
In contrast, our method consistently reduces this disruption.

To provide an intuitive illustration of the disruptions, we summarize two representative cases observed in our analysis. 
For a user with a long history of purchasing sensitive-skin products, given the query ``Lightweight foundation for sensitive, acne-prone skin'', the PEFT baseline may prioritize popular but fragranced foundations, effectively relaxing the sensitivity constraints, whereas GEMS places products tailored to sensitive skin at the top, better adhering to the query-specific requirements. 
Similarly, for a user whose long-term preference is consistently small-dog health products but who recently clicked generic pet items, PEFT can drift toward cat-related recommendations, while GEMS remains aligned with the stable profile and recommends small-dog supplements and related dog products. 
Overall, these examples reflect that PEFT updates can bias the model toward spurious or short-term signals and weaken constraint-sensitive intent reasoning, whereas GEMS better preserves such intent interpretation under unified S\&R tuning.


\begin{table}[t]
{\small \centering
\caption{Efficiency comparison between our method, LoRA, and LoRA-MoE under $W \in \mathbb{R}^{m \times n}$, rank $r$, $E$ LoRA experts in LoRA-MOE.}
\label{tab:efficiency}
\begin{tabular}{lccc}
\toprule
 & \textbf{Ours} & \textbf{LoRA} & \textbf{LoRA-MoE} \\
\midrule
Weights & $mn$ & $mn + mr + nr$ & $mn + E(mr + nr)$ \\
Optimizer States & $2mr + 4nr$ & $2mr + 2nr$ & $2E(mr + nr)$ \\
Multi-Subspace & \checkmark & $\times$ & $\times$ \\
Pre-Training & \checkmark & $\times$ & $\times$ \\
Fine-Tuning & \checkmark & \checkmark & \checkmark \\
\bottomrule
\end{tabular}}
\end{table}

\begin{figure}
  \centering 
  \subfigure{
    \includegraphics[width=1.6in]{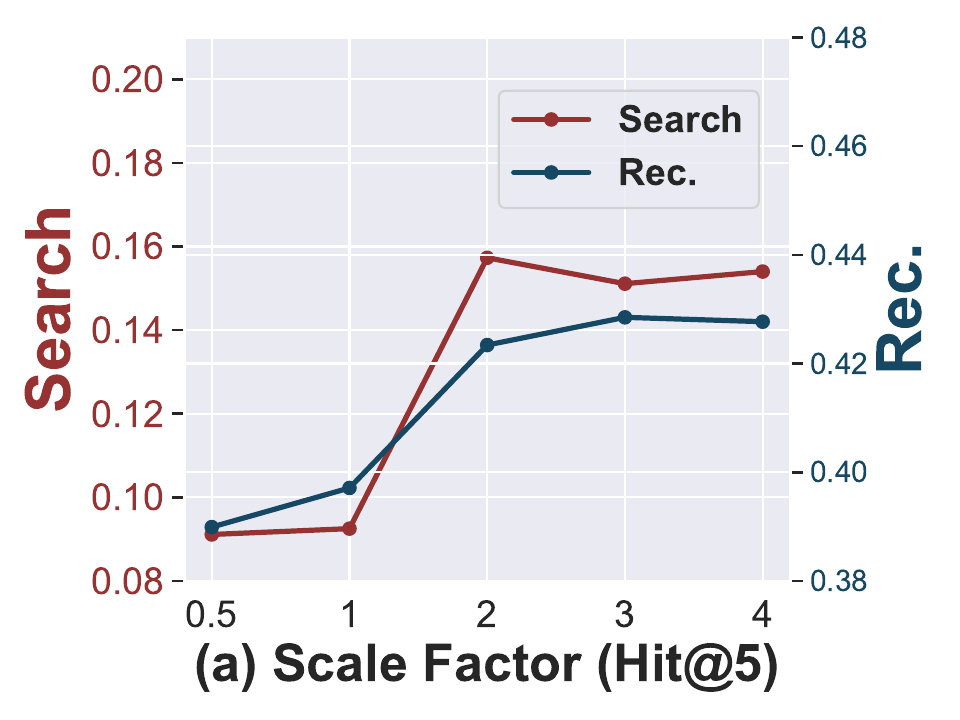}} 
  \subfigure{
     \includegraphics[width=1.6in]{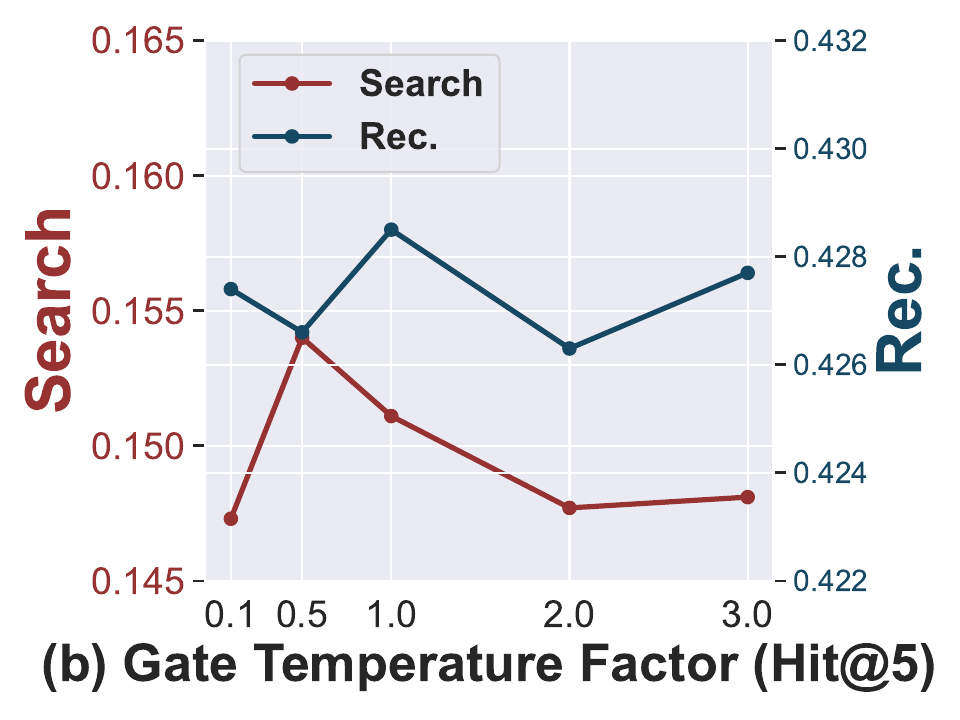}} 
\caption{Hyperparameter analysis on Qilin.}
  \label{fig:hyper}
\end{figure}

\subsubsection{Efficiency analysis} 
\label{sec:eff}
We analyze the efficiency of our method compared with representative PEFT methods: LoRA~\cite{hu2022lora} and its MoE-style variants~\cite{dou2023loramoe}.
Let a linear transformation layer in LLMs have a weight matrix $W \in \mathbb{R}^{m \times n}$ with $m \leq n$.
The memory costs can be decomposed into two parts: 
(1) \emph{weights} that must be stored and accessed during both training and inference, and 
(2) \emph{training-time states} that are only needed during optimization.
The latter includes optimizer moments and, for our methods, the projection bases used to form low-dimensional updates.

For weights, GEMS does not introduce any additional adapter weights beyond the original parameter matrix, resulting in a cost of $mn$. 
In contrast, LoRA introduces two low-rank matrices $A \in \mathbb{R}^{m \times r}$ and $B \in \mathbb{R}^{r \times n}$, requiring an additional $mr + nr$ parameters on top of $W$.
Similarly, LoRA-MoE with $E$ experts stores $E$ such adapters, i.e., $E(mr+nr)$ extra weights (router parameters omitted).
For training-time states, our method maintains the projection bases and first- and second-order momentum estimates for three subspaces: one shared subspace of dimension $r$, and two task-specific subspaces of dimension $r/2$ each. 
This yields a total state size of 
\[
(mr + 2nr) + (mr/2 + nr) + (mr/2 + nr) = 2mr + 4nr,
\]
where each term corresponds to one subspace (projection basis plus two moment estimates).
By contrast, LoRA requires optimizer states for both $A$ and $B$, leading to a total of $2mr + 2nr$; LoRA-MoE scales accordingly with the number of experts.

Table~\ref{tab:efficiency} summarizes the comparison. 
Although our number of training-time states is larger than LoRA due to the multi-subspace design, it is often smaller than LoRA-MoE when multiple experts are used (e.g., $E\!\ge\!2$).
Additionally, our method achieves two clear advantages: 
(i) no additional weight parameters are introduced, which substantially reduces long-term storage and deployment overhead, and 
(ii) GEMS can naturally support both pre-training and fine-tuning, since it operates directly on the original parameter matrix $W$ without introducing separate adaptation modules, whereas LoRA-style methods are limited to fine-tuning.
Since inference and deployment are dominated by served weights, GEMS is more deployment-efficient for large-scale unified S\&R.

\subsubsection{Hyperparameter analysis (RQ4)} 
\label{subsec:hyperparameters}
We analyze two critical hyperparameters: the scale factor $\alpha$ and the gate temperature factor $\tau$. 
This analysis is performed on T5-base, presented in Figure~\ref{fig:hyper}. 
Our findings are as follows:
(1) \textbf{Effect of $\alpha$.} 
The scale factor $\alpha$ controls the strength of the low-rank update.
When $\alpha$ is too small, insufficient task signals are injected, leading to under-updating and suboptimal performance on both S\&R.
Conversely, an excessively large $\alpha$ degrades performance, likely due to unstable training dynamics such as overshooting or diverging from optimal basins.
(2) \textbf{Effect of $\tau$.} 
The temperature $\tau$ controls the sharpness of adaptive weighting.
Very small $\tau$ yields near one-hot gating, suppressing one task, while very large $\tau$ degenerates to uniform weights. 
Intermediate $\tau$ achieves robust trade-offs, adaptively balancing S\&R. 
(3) \textbf{Task-specific Optima}.
We observe slight shifts in the optimal hyperparameter ranges between S\&R, reflecting their inherent task-specific characteristics.
However, there exist broad overlapping regions where both tasks achieve strong performance, indicating that our framework is generally robust to hyperparameter settings.

\vfill\eject
\section{Conclusion}
\label{sec:conclusion}
We proposed GEMS, a Gradient Multi-Subspace Tuning framework that unifies search and recommendation. 
By projecting gradients into low-rank subspaces and mapping them back to the full parameter space, GEMS reduces training-time memory overhead while enabling effective optimization. 
Furthermore, GEMS comprises Multi-Subspace Decomposition to mitigate gradient conflicts and Null-Space Projection to preserve general-domain knowledge. 
Across benchmark S\&R tasks, GEMS outperforms state-of-the-art baselines, reducing gradient conflicts and semantic drift.

Our findings suggest a promising direction for jointly optimizing search and recommendation within a single LLM. 
However, GEMS currently preserves general-domain knowledge holistically and relies on fixed subspace ranks, limiting selective retention and adaptive capacity. 
Future work will explore selective knowledge preservation for S\&R and adaptive subspace learning with dynamically adjusted ranks.

\clearpage

{
\tiny
\bibliographystyle{ACM-Reference-Format}
\balance
\bibliography{bibfile}

@article{du2018adapting,
  title={Adapting auxiliary losses using gradient similarity},
  author={Du, Yunshu and Czarnecki, Wojciech M and Jayakumar, Siddhant M and Farajtabar, Mehrdad and Pascanu, Razvan and Lakshminarayanan, Balaji},
  journal={arXiv:1812.02224},
  year={2018}
}

@inproceedings{He2017Neural,
  title = {Neural collaborative filtering},
  author = {He, Xiangnan and Liao, Lizi and Zhang, Hanwang and Nie, Liqiang and Hu, Xia and Chua, Tat-Seng},
  booktitle={WWW},
  pages = {173-182},
  year={2017},
  publisher={ACM}
}

@inproceedings{kingma2014adam,
  author       = {Diederik P. Kingma and
                  Jimmy Ba},
  title        = {Adam: {A} Method for Stochastic Optimization},
  booktitle    = {ICLR},
  year         = {2015},
}

@inproceedings{sun2023learning,
  author       = {Weiwei Sun and
                  Lingyong Yan and
                  Zheng Chen and
                  Shuaiqiang Wang and
                  Haichao Zhu and
                  Pengjie Ren and
                  Zhumin Chen and
                  Dawei Yin and
                  Maarten de Rijke and
                  Zhaochun Ren},
  title        = {Learning to Tokenize for Generative Retrieval},
  booktitle    = {NeurIPS},
  year         = {2023},
}

@article{shi2025unified,
  title={Unified Generative Search and Recommendation},
  author={Shi, Teng and Xu, Jun and Zhang, Xiao and Zang, Xiaoxue and Zheng, Kai and Song, Yang and Yu, Enyun},
  journal={arXiv:2504.05730},
  year={2025}
}

@inproceedings{si2023search,
  title={When search meets recommendation: Learning disentangled search representation for recommendation},
  author={Si, Zihua and Sun, Zhongxiang and Zhang, Xiao and Xu, Jun and Zang, Xiaoxue and Song, Yang and Gai, Kun and Wen, Ji-Rong},
  booktitle={SIGIR},
  publisher={ACM},
  pages={1313--1323},
  year={2023}
}

@inproceedings{shi2024unisar,
  title={UniSAR: Modeling User Transition Behaviors between Search and Recommendation},
  author={Shi, Teng and Si, Zihua and Xu, Jun and Zhang, Xiao and Zang, Xiaoxue and Zheng, Kai and Leng, Dewei and Niu, Yanan and Song, Yang},
  booktitle={SIGIR},
  publisher={ACM},
  pages={1029--1039},
  year={2024}
}

@inproceedings{ai2017learning,
  title={Learning a hierarchical embedding model for personalized product search},
  author={Ai, Qingyao and Zhang, Yongfeng and Bi, Keping and Chen, Xu and Croft, W Bruce},
  booktitle={SIGIR},
  publisher={ACM},
  pages={645--654},
  year={2017}
}

@inproceedings{zamani2018joint,
  author       = {Hamed Zamani and
                  W. Bruce Croft},
  title        = {Joint Modeling and Optimization of Search and Recommendation},
  booktitle    = {DESIRE},
  volume       = {2167},
  pages        = {36--41},
  publisher    = {CEUR-WS.org},
  year         = {2018},
}

@inproceedings{yao2021user,
  title={USER: A unified information search and recommendation model based on integrated behavior sequence},
  author={Yao, Jing and Dou, Zhicheng and Xie, Ruobing and Lu, Yanxiong and Wang, Zhiping and Wen, Ji-Rong},
  booktitle={CIKM},
  publisher={ACM},
  pages={2373--2382},
  year={2021}
}

@inproceedings{xie2024unifiedssr,
  title={UnifiedSSR: A Unified Framework of Sequential Search and Recommendation},
  author={Xie, Jiayi and Liu, Shang and Cong, Gao and Chen, Zhenzhong},
  booktitle={WWW},
  pages={3410--3419},
  publisher={ACM},
  year={2024}
}

@inproceedings{rajput2023recommender,
  author       = {Shashank Rajput and
                  Nikhil Mehta and
                  Anima Singh and
                  Raghunandan Hulikal Keshavan and
                  Trung Vu and
                  Lukasz Heldt and
                  Lichan Hong and
                  Yi Tay and
                  Vinh Q. Tran and
                  Jonah Samost and
                  Maciej Kula and
                  Ed H. Chi and
                  Mahesh Sathiamoorthy},
  title        = {Recommender Systems with Generative Retrieval},
  booktitle    = {NeurIPS},
  year         = {2023},
}

@article{wang2024learnable,
  title={Learnable Tokenizer for LLM-based Generative Recommendation},
  author={Wang, Wenjie and Bao, Honghui and Lin, Xinyu and Zhang, Jizhi and Li, Yongqi and Feng, Fuli and Ng, See-Kiong and Chua, Tat-Seng},
  journal={arXiv:2405.07314},
  year={2024}
}

@inproceedings{penha2024bridging,
  title={Bridging Search and Recommendation in Generative Retrieval: Does One Task Help the Other?},
  author={Penha, Gustavo and Vardasbi, Ali and Palumbo, Enrico and De Nadai, Marco and Bouchard, Hugues},
  booktitle={RecSys},
  publisher={ACM},
  pages={340--349},
  year={2024}
}

@inproceedings{li2024learning,
  title={Learning to rank in generative retrieval},
  author={Li, Yongqi and Yang, Nan and Wang, Liang and Wei, Furu and Li, Wenjie},
  booktitle={AAAI},
  publisher={AAAI press},
  pages={8716--8723},
  year={2024}
}

@inproceedings{zhao2025model,
  title={Model Meets Knowledge: Analyzing Knowledge Types for Conversational Recommender Systems},
  author={Zhao, Jujia and Wang, Yumeng and Ren, Zhaochun and Verberne, Suzan},
  booktitle={Proceedings of the Nineteenth ACM Conference on Recommender Systems},
  pages={802--811},
  year={2025}
}

@article{lin2026verifiable,
  title={Verifiable Reasoning for LLM-based Generative Recommendation},
  author={Lin, Xinyu and Zeng, Hanqing and Yu, Hanchao and Xia, Yinglong and Zhang, Jiang and Singh, Aashu and Liu, Fei and Wang, Wenjie and Feng, Fuli and Chua, Tat-Seng and others},
  journal={arXiv preprint arXiv:2603.07725},
  year={2026}
}

@inproceedings{lin2026bringing,
  title={Bringing Reasoning to Generative Recommendation Through the Lens of Cascaded Ranking},
  author={Lin, Xinyu and Liu, Pengyuan and Wang, Wenjie and Hu, Yicheng and Xu, Chen and Feng, Fuli and Wang, Qifan and Chua, Tat-Seng},
  booktitle={Proceedings of the ACM Web Conference 2026},
  pages={8939--8949},
  year={2026}
}

@inproceedings{xu2026unveiling,
  title={Unveiling and Simulating Short-Video Addiction Behaviors via Economic Addiction Theory},
  author={Xu, Chen and Yi, Zhipeng and Wang, Ruizi and Wang, Wenjie and Xu, Jun and de Rijke, Maarten},
  booktitle={Proceedings of the ACM Web Conference 2026},
  pages={5987--5997},
  year={2026}
}

@inproceedings{zhang2024towards,
  title={Towards empathetic conversational recommender systems},
  author={Zhang, Xiaoyu and Xie, Ruobing and Lyu, Yougang and Xin, Xin and Ren, Pengjie and Liang, Mingfei and Zhang, Bo and Kang, Zhanhui and de Rijke, Maarten and Ren, Zhaochun},
  booktitle={RecSys},
  pages={84--93},
  publisher={ACM},
  year={2024}
}

@inproceedings{wu2024generative,
  title={Generative retrieval as multi-vector dense retrieval},
  author={Wu, Shiguang and Wei, Wenda and Zhang, Mengqi and Chen, Zhumin and Ma, Jun and Ren, Zhaochun and de Rijke, Maarten and Ren, Pengjie},
  booktitle={SIGIR},
  publisher={ACM},
  pages={1828--1838},
  year={2024}
}

@inproceedings{hu2022lora,
  author       = {Edward J. Hu and
                  Yelong Shen and
                  Phillip Wallis and
                  Zeyuan Allen{-}Zhu and
                  Yuanzhi Li and
                  Shean Wang and
                  Lu Wang and
                  Weizhu Chen},
  title        = {LoRA: Low-Rank Adaptation of Large Language Models},
  booktitle    = {ICLR},
  publisher    = {OpenReview.net},
  year         = {2022},}

@inproceedings{penha2025semantic,
  title={Semantic IDs for Joint Generative Search and Recommendation},
  author={Penha, Gustavo and D'Amico, Edoardo and De Nadai, Marco and Palumbo, Enrico and Tamborrino, Alexandre and Vardasbi, Ali and Lefarov, Max and Lin, Shawn and Heath, Timothy and Fabbri, Francesco and others},
  booktitle={RecSys},
  pages={1296--1301},
  year={2025}
}

@inproceedings{fang2024alphaedit,
  author       = {Junfeng Fang and
                  Houcheng Jiang and
                  Kun Wang and
                  Yunshan Ma and
                  Jie Shi and
                  Xiang Wang and
                  Xiangnan He and
                  Tat{-}Seng Chua},
  title        = {AlphaEdit: Null-Space Constrained Knowledge Editing for Language Models},
  booktitle    = {ICLR},
  publisher    = {OpenReview.net},
  year         = {2025},
}

@inproceedings{zhao2024galore,
  author       = {Jiawei Zhao and
                  Zhenyu Zhang and
                  Beidi Chen and
                  Zhangyang Wang and
                  Anima Anandkumar and
                  Yuandong Tian},
  title        = {GaLore: Memory-Efficient {LLM} Training by Gradient Low-Rank Projection},
  booktitle    = {ICML},
  publisher    = {OpenReview.net},
  year         = {2024},
}

@article{xia2024assessing,
  title={Assessing parameter efficient methods for pre-trained language model in annotating scRNA-seq data},
  author={Xia, Yucheng and Liu, Yuhang and Li, Tianhao and He, Sihan and Chang, Hong and Wang, Yaqing and Zhang, Yongqing and Ge, Wenyi},
  journal={Methods},
  volume={228},
  pages={12--21},
  year={2024},
  publisher={Elsevier}
}

@article{zhao2025unifying,
  title={Unifying Search and Recommendation with Dual-View Representation Learning in a Generative Paradigm},
  author={Zhao, Jujia and Wang, Wenjie and Xu, Chen and Chen, Xiuying and Ren, Zhaochun and Verberne, Suzan},
  journal={arXiv:2504.06714},
  year={2025}
}

@article{dou2023loramoe,
  title={Loramoe: Revolutionizing mixture of experts for maintaining world knowledge in language model alignment},
  author={Dou, Shihan and Zhou, Enyu and Liu, Yan and Gao, Songyang and Zhao, Jun and Shen, Wei and Zhou, Yuhao and Xi, Zhiheng and Wang, Xiao and Fan, Xiaoran and others},
  journal={arXiv:2312.09979},
  volume={4},
  number={7},
  year={2023}
}

@article{xiong2020approximate,
  title={Approximate nearest neighbor negative contrastive learning for dense text retrieval},
  author={Xiong, Lee and Xiong, Chenyan and Li, Ye and Tang, Kwok-Fung and Liu, Jialin and Bennett, Paul and Ahmed, Junaid and Overwijk, Arnold},
  journal={arXiv:2007.00808},
  year={2020}
}

@article{zhou2023webultron,
  title={WebUltron: An ultimate retriever on webpages under the model-centric paradigm},
  author={Zhou, Yujia and Yao, Jing and Wu, Ledell and Dou, Zhicheng and Wen, Ji-Rong},
  journal={IEEE Transactions on Knowledge and Data Engineering},
  volume={36},
  number={9},
  pages={4996--5006},
  year={2023},
  publisher={IEEE}
}

@inproceedings{sanh2021multitask,
  author       = {Victor Sanh and
                  Albert Webson and
                  Colin Raffel and
                  Stephen H. Bach and
                  Lintang Sutawika and
                  Zaid Alyafeai and
                  Antoine Chaffin and
                  Arnaud Stiegler and
                  Arun Raja and
                  Manan Dey and
                  M Saiful Bari and
                  Canwen Xu and
                  Urmish Thakker and
                  Shanya Sharma Sharma and
                  Eliza Szczechla and
                  Taewoon Kim and
                  Gunjan Chhablani and
                  Nihal V. Nayak and
                  Debajyoti Datta and
                  Jonathan Chang and
                  Mike Tian{-}Jian Jiang and
                  Han Wang and
                  Matteo Manica and
                  Sheng Shen and
                  Zheng Xin Yong and
                  Harshit Pandey and
                  Rachel Bawden and
                  Thomas Wang and
                  Trishala Neeraj and
                  Jos Rozen and
                  Abheesht Sharma and
                  Andrea Santilli and
                  Thibault F{\'{e}}vry and
                  Jason Alan Fries and
                  Ryan Teehan and
                  Teven Le Scao and
                  Stella Biderman and
                  Leo Gao and
                  Thomas Wolf and
                  Alexander M. Rush},
  title        = {Multitask Prompted Training Enables Zero-Shot Task Generalization},
  booktitle    = {ICLR},
  publisher    = {OpenReview.net},
  year         = {2022},
}

@inproceedings{wei2021finetuned,
  author       = {Jason Wei and
                  Maarten Bosma and
                  Vincent Y. Zhao and
                  Kelvin Guu and
                  Adams Wei Yu and
                  Brian Lester and
                  Nan Du and
                  Andrew M. Dai and
                  Quoc V. Le},
  title        = {Finetuned Language Models are Zero-Shot Learners},
  booktitle    = {ICLR},
  year         = {2022},
}

@inproceedings{huang2024unifit,
  title={UNIFIT: A Unified Framework For Instruction Tuning To Improve Instruction Following Ability For Large Language Models},
  author={Huang, Qiang and Huang, Feng and Tao, DeHao and Wang, BingKun and Huang, YongFeng},
  booktitle={Proceedings of the Annual Meeting of the Cognitive Science Society},
  volume={46},
  year={2024}
}

@article{shengyu2023instruction,
  title={Instruction tuning for large language models: A survey},
  author={Shengyu, Zhang and Linfeng, Dong and Xiaoya, Li and Sen, Zhang and Xiaofei, Sun and Shuhe, Wang and Jiwei, Li and Hu, Runyi and Tianwei, Zhang and Wu, Fei and others},
  journal={arXiv:2308.10792},
  year={2023}
}

@inproceedings{pfeiffer2020adapterfusion,
  author       = {Jonas Pfeiffer and
                  Aishwarya Kamath and
                  Andreas R{\"{u}}ckl{\'{e}} and
                  Kyunghyun Cho and
                  Iryna Gurevych},
  title        = {AdapterFusion: Non-Destructive Task Composition for Transfer Learning},
  booktitle    = {EACL},
  pages        = {487--503},
  publisher    = {ACL},
  year         = {2021},
}

@inproceedings{ding2023mitigating,
  title={Mitigating task interference in multi-task learning via explicit task routing with non-learnable primitives},
  author={Ding, Chuntao and Lu, Zhichao and Wang, Shangguang and Cheng, Ran and Boddeti, Vishnu Naresh},
  booktitle={CVPR},
  pages={7756--7765},
  publisher = {{IEEE}},
  year={2023}
}

@article{chung2024scaling,
  title={Scaling instruction-finetuned language models},
  author={Chung, Hyung Won and Hou, Le and Longpre, Shayne and Zoph, Barret and Tay, Yi and Fedus, William and Li, Yunxuan and Wang, Xuezhi and Dehghani, Mostafa and Brahma, Siddhartha and others},
  journal={Journal of Machine Learning Research},
  volume={25},
  number={70},
  pages={1--53},
  year={2024}
}

@article{raffel2020exploring,
  title={Exploring the limits of transfer learning with a unified text-to-text transformer},
  author={Raffel, Colin and Shazeer, Noam and Roberts, Adam and Lee, Katherine and Narang, Sharan and Matena, Michael and Zhou, Yanqi and Li, Wei and Liu, Peter J},
  journal={Journal of machine learning research},
  volume={21},
  number={140},
  pages={1--67},
  year={2020}
}

@article{fedus2022switch,
  title={Switch transformers: Scaling to trillion parameter models with simple and efficient sparsity},
  author={Fedus, William and Zoph, Barret and Shazeer, Noam},
  journal={Journal of Machine Learning Research},
  volume={23},
  number={120},
  pages={1--39},
  year={2022}
}

@article{lepikhin2020gshard,
  title={Gshard: Scaling giant models with conditional computation and automatic sharding},
  author={Lepikhin, Dmitry and Lee, HyoukJoong and Xu, Yuanzhong and Chen, Dehao and Firat, Orhan and Huang, Yanping and Krikun, Maxim and Shazeer, Noam and Chen, Zhifeng},
  journal={arXiv:2006.16668},
  year={2020}
}

@inproceedings{zhang2024unified,
  title={Unified Dual-Intent Translation for Joint Modeling of Search and Recommendation},
  author={Zhang, Yuting and Wu, Yiqing and Han, Ruidong and Sun, Ying and Zhu, Yongchun and Li, Xiang and Lin, Wei and Zhuang, Fuzhen and An, Zhulin and Xu, Yongjun},
 publisher    = {{ACM}},
  booktitle={KDD},
  pages={6291--6300},
  year={2024}

}

@inproceedings{zhao2022joint,
  title={Joint learning of e-commerce search and recommendation with a unified graph neural network},
  author={Zhao, Kai and Zheng, Yukun and Zhuang, Tao and Li, Xiang and Zeng, Xiaoyi},
  booktitle={WSDM},
  pages={1461--1469},
  year={2022}
}

@inproceedings{wang2022super,
  author       = {Yizhong Wang and
                  Swaroop Mishra and
                  Pegah Alipoormolabashi and
                  Yeganeh Kordi and
                  Amirreza Mirzaei and
                  Atharva Naik and
                  Arjun Ashok and
                  Arut Selvan Dhanasekaran and
                  Anjana Arunkumar and
                  David Stap and
                  Eshaan Pathak and
                  Giannis Karamanolakis and
                  Haizhi Gary Lai and
                  Ishan Purohit and
                  Ishani Mondal and
                  Jacob Anderson and
                  Kirby Kuznia and
                  Krima Doshi and
                  Kuntal Kumar Pal and
                  Maitreya Patel and
                  Mehrad Moradshahi and
                  Mihir Parmar and
                  Mirali Purohit and
                  Neeraj Varshney and
                  Phani Rohitha Kaza and
                  Pulkit Verma and
                  Ravsehaj Singh Puri and
                  Rushang Karia and
                  Savan Doshi and
                  Shailaja Keyur Sampat and
                  Siddhartha Mishra and
                  Sujan Reddy A and
                  Sumanta Patro and
                  Tanay Dixit and
                  Xudong Shen},
  title        = {Super-NaturalInstructions: Generalization via Declarative Instructions
                  on 1600+ {NLP} Tasks},
  booktitle    = {EMNLP},
  pages        = {5085--5109},
  publisher    = {ACL},
  year         = {2022},}

@inproceedings{feng2024mixture,
  author       = {Wenfeng Feng and
                  Chuzhan Hao and
                  Yuewei Zhang and
                  Yu Han and
                  Hao Wang},
  title        = {Mixture-of-LoRAs: An Efficient Multitask Tuning Method for Large Language
                  Models},
  booktitle    = {LREC/COLING},
  pages        = {11371--11380},
  publisher    = {{ELRA} and {ICCL}},
  year         = {2024},
}

@inproceedings{yang2025mtl,
  title={Mtl-lora: Low-rank adaptation for multi-task learning},
  author={Yang, Yaming and Muhtar, Dilxat and Shen, Yelong and Zhan, Yuefeng and Liu, Jianfeng and Wang, Yujing and Sun, Hao and Deng, Weiwei and Sun, Feng and Zhang, Qi and others},
  booktitle={AAAI},
  volume={39},
  number={20},
  pages={22010--22018},
  year={2025}
}

@article{wei2022emergent,
  author       = {Jason Wei and
                  Yi Tay and
                  Rishi Bommasani and
                  Colin Raffel and
                  Barret Zoph and
                  Sebastian Borgeaud and
                  Dani Yogatama and
                  Maarten Bosma and
                  Denny Zhou and
                  Donald Metzler and
                  Ed H. Chi and
                  Tatsunori Hashimoto and
                  Oriol Vinyals and
                  Percy Liang and
                  Jeff Dean and
                  William Fedus},
  title        = {Emergent Abilities of Large Language Models},
  journal      = {Transactions on Machine Learning Research},
  volume       = {2022},
  year         = {2022},
}

@inproceedings{zhang2026model,
  title={Model Editing for New Document Integration in Generative Information Retrieval},
  author={Zhang, Zhen and Wang, Zihan and Ma, Xinyu and Wang, Shuaiqiang and Yin, Dawei and Xin, Xin and Ren, Pengjie and de Rijke, Maarten and Ren, Zhaochun},
  booktitle={WWW},
  pages={1993--2003},
  year={2026}
}

@article{wang2025graph,
  title={Graph-enhanced prompt learning for cross-domain contract element extraction},
  author={Wang, Zihan and Wang, Hanbing and Ren, Pengjie and Chen, Zhumin and De Rijke, Maarten and Ren, Zhaochun},
  journal={ACM Transactions on Information Systems},
  volume={43},
  number={3},
  pages={1--24},
  year={2025},
}

@inproceedings{wang2025cooperative,
  title={A cooperative multi-agent framework for zero-shot named entity recognition},
  author={Wang, Zihan and Zhao, Ziqi and Lyu, Yougang and Chen, Zhumin and de Rijke, Maarten and Ren, Zhaochun},
  booktitle={WWW},
  pages={4183--4195},
  year={2025}
}

@article{lyu2024macpo,
  title={Macpo: Weak-to-strong alignment via multi-agent contrastive preference optimization},
  author={Lyu, Yougang and Yan, Lingyong and Wang, Zihan and Yin, Dawei and Ren, Pengjie and de Rijke, Maarten and Ren, Zhaochun},
  journal={arXiv:2410.07672},
  year={2024}
}
}

\clearpage


\end{document}